\documentclass[%
 reprint,
superscriptaddress,
nofootinbib,
 amsmath,amssymb,
 aps,
]{revtex4-2}

\usepackage{graphicx}
\usepackage{dcolumn}
\usepackage{bm}
\usepackage{braket}
\usepackage{mathtools}

\newcommand{\appropto}{\mathrel{\vcenter{
  \offinterlineskip\halign{\hfil$##$\cr
    \propto\cr\noalign{\kern2pt}\sim\cr\noalign{\kern-2pt}}}}}

\begin{document}

\preprint{WORKING PAPER}

\title{How Democracies Polarize: A Multilevel Perspective} 

\author{Sihao Huang}
\affiliation{Department of Physics, Massachusetts Institute of Technology, Cambridge, MA, USA}
\affiliation{Center for Constructive Communication, MIT Media Lab, Cambridge, MA, USA}
\author{Alexander F. Siegenfeld}
\affiliation{Department of Physics, Massachusetts Institute of Technology, Cambridge, MA, USA}
\affiliation{Center for Constructive Communication, MIT Media Lab, Cambridge, MA, USA}
\author{Andrew Gelman}
\affiliation{Departments of Statistics and Political Science, Columbia University, New York, NY, USA}


\begin{abstract}

Democracies employ elections at various scales to select officials at the corresponding levels of administration. The geographic distribution of political opinion, the policy issues delegated to each level, and the multilevel interactions between elections can all greatly impact the makeup of these representative bodies. This perspective is not new: the adoption of federal systems has been motivated by the idea that they possess desirable traits not provided by democracies on a single scale. Yet most existing models of polarization do not capture how nested local and national elections interact with heterogeneous political geographies. We begin by developing a framework to describe the multilevel distribution of opinions and analyze the flow of variance among geographic scales, applying it to historical data in the United States from 1912 to 2020. We describe how unstable elections can arise due to the spatial distribution of opinions and how tradeoffs occur between national and local elections. We also examine multi-dimensional spaces of political opinion, for which we show that a decrease in local salience can constrain the dimensions along which elections occur, preventing a federal system from serving as an effective safeguard against polarization. These analyses, based on the interactions between elections and opinion distributions at various scales, offer insights into how democracies can be strengthened to mitigate polarization and increase electoral representation.

\end{abstract}

\maketitle

\section{Introduction}
\label{sec:introduction}

Polities often span vast geographic regions and encompass groups with diverse interests. Democracies contend with this heterogeneity by distributing representation and governance across levels. Since at least the eighteenth century, political philosophers have emphasized the need to balance collective action with citizen representation, with Montesquieu arguing that multilevel governance combines ``the internal advantages of a [small] republic" with the ``external force of a monarchical government" \cite{montesquieu_spirit_2002}. Other scholars like James Madison also believed that decentralized governance could guard against polarization. Madison asserted in Federalist No. 10 that a ``pure democracy \dots can admit of no cure for the mischief of factions," while a well constructed Union has a ``tendency to break and control the violence of faction" \cite{madison_federalist_1787}. One of their key contentions was that democracies must be built as multiscale systems, motivated by the heterogeneous distribution of voters and the need to devolve policy-making responsibilities. This advocacy was taken up with great enthusiasm: by the turn of the 21st century, 95\% of all democracies were electing subnational tiers of government \cite{yusuf_decentralization_2000}.

How this multilevel system interacts with political geography can greatly impact the type of polarization observed in a country. Even if the overall set of voter opinions is fixed, considerably different electoral outcomes can result depending on the geographic distributions of these opinions \cite{kendall_law_1950, gelman_unified_1994, mccarty_geography_2019}. For a hypothetical nation in which differing opinions are geographically well-mixed, most political contention would be resolved locally while larger-scale politics remains depolarized. On the other extreme, in a nation in which voter opinions are perfectly sorted into districts, all contention must be resolved through larger-scale elections or legislative bodies. Real democracies lie between these extremes, with opinion variance spread across levels. Disagreement is resolved in a multilevel fashion, with some compromise achieved through local government or the election of legislators from localized districts and some through larger-scale elections and legislative bodies. 

Put mathematically, democratic governance at any particular scale is a mean-field treatment\footnote{See section III of reference \cite{intro_to_complex_systems2020} for a non-technical review of mean-field theory and the conditions under which it applies.} of the electorate, in that the disparate views and needs must be collapsed into a single instrument that works on average. However, the distribution of political opinions is often poorly described by a mean-field theory due to strong geographic correlations across multiple spatial scales, born out of factors such as urban history, clustering, and social ties \cite{rodden_geographic_2010, gelman_mathematics_2002}. For instance, models that assume no geographic correlations between voters yield behaviors governed by the central limit theorem, which have been empirically shown to overestimate the probability of close elections in larger jurisdictions. These inappropriate assumptions have misleading political implications regarding how jurisdictions should be weighted in bodies such as the Electoral College \cite{gelman_standard_2004}. 

Furthermore, such correlations in political opinion suggest a potential mismatch between the electorate and the mean-field instruments---which are best suited for systems in which deviations from the mean are sufficiently uncorrelated---that would represent it. This mismatch can lead to failures in the electoral process, which include \textit{unstable} elections---in which a slight change in electoral opinions can lead to a large swing in the election outcome---and \textit{negative representation}---in which a shift in one's opinion position can move the outcome in the opposite direction \cite{siegenfeld_bar-yam_2020}. Democracies, therefore, need to take into consideration not only how the geographic distribution of opinions affects the fairness of multistage elections (e.g., via districting and apportionment) \cite{gelman_mathematics_2002, preference_aggregation}, but also the makeup of representative bodies and the devolution of policy scope across scales. 

Nationalization of political discourse further aggravates the discrepancy between how institutions are designed and how political opinions are distributed. The United States, for instance, has seen a decline in spatially-bound media, a deepening urban-rural divide, and further centralization of government authority in recent decades. Gubernatorial elections have become increasingly aligned with state-level presidential votes since the 1980s and are almost perfectly predicted today using presidential ballots in those districts without state-specific information \cite{hopkins_increasingly_2018}. The original architects of America's federal system had assumed that the ``first and most natural attachment of the people will be to the governments of their respective states" \cite{madison_federalist_1788}, but this premise has been gradually eroded. Campaign contributions, voter turnout, and search interests indicate that Americans identify overwhelmingly with national party politics, undermining the ability of the federal system to guard against polarization.  

Starting with a framework that takes the distribution of political opinions as given (and therefore applies regardless of the mechanism of opinion formation), we derive a model for understanding polarization and representation in multilevel democracies. We first provide a mathematical formalism for describing opinion heterogeneity in section~\ref{sec:coarse-graining} and examine how this geographic heterogeneity affects elections in the context of social ties and segregation in section~\ref{sec:socialization}. We connect this formalism to empirical data by analyzing how the spread of opinion variance across spatial scales has changed over time in the United States and how these changes relate to the level of polarization in representative bodies.

The analysis thus far applies to electoral systems with any number of parties or active issue dimensions. However, phenomena such as the ideological alignment between local and national elections require us to explicitly consider multiple issue dimensions \cite{oates_fiscal_1972}. In section~\ref{sec:multidimensional}, we explore the consequences of opinion measurements and introduce the concept of an \textit{election subspace}. This multidimensional analysis offers an explanation for why measurements of mass polarization often lag behind elite polarization and reveals that elections have a tendency to occur along the axis of maximum opinion variance. In a system with multiple levels, a tradeoff emerges between national and local elections, resulting in higher electoral variance for elections that play a bigger role in defining political discourse. Combining these arguments with an analysis of social interactions and the geographic distribution of opinions leads to the conclusion that greater national (as opposed to local) salience leads to increased polarization and instability in larger-scale elections. These results parallel the situation in the United States, in which ``hollowed-out," ``top-heavy" parties that used to be largely local have led to increasingly unstable national elections and non-competitive local offices \cite{lee_hollow_2019}. 

Scholarship on fiscal federalism has shown that ``not all federations are created equal" \cite{rodden_hamiltons_2006}. Different relationships between the political composition and fiscal structures of various level of government can lead to vastly divergent financial outcomes. Each level of government also has its comparative advantages. For instance, the need for tight feedback and diseconomies of scale may make local governments more effective implementers of developmental policy, while the requirement to coordinate regional policies that prevent a race to the bottom can make the national government more suited to redistribution \cite{peterson_price_1995}. 

We suggest that in addition to matching the multilevel complexity~\cite{intro_to_complex_systems2020} of government to that of the policy environment, the multilevel distribution of opinions must also be considered. In other words, the efficacy of a federal democracy rests on three pillars: the multilevel structure of political institutions, the multilevel complexity of the policy environment, and the multilevel distribution of political opinions. The first two pillars cannot be considered independently of the third if citizen opinions are to be well represented. Apart from normative concerns, a failure to stably represent these opinions can result in gridlock and extreme levels of polarization. Our analysis finds a tradeoff between larger-scale governance and the satisfaction of citizen preferences---the variance of which increases with geographic scale. Devolving powers to local levels can reduce negative representation and electoral instability, particularly if such powers are along issue dimensions for which there is substantial geographic polarization and segregation. 

\section{How differences in opinion are distributed across geographic scales}
\label{sec:coarse-graining}

\begin{figure*}
\includegraphics[width=1\textwidth]{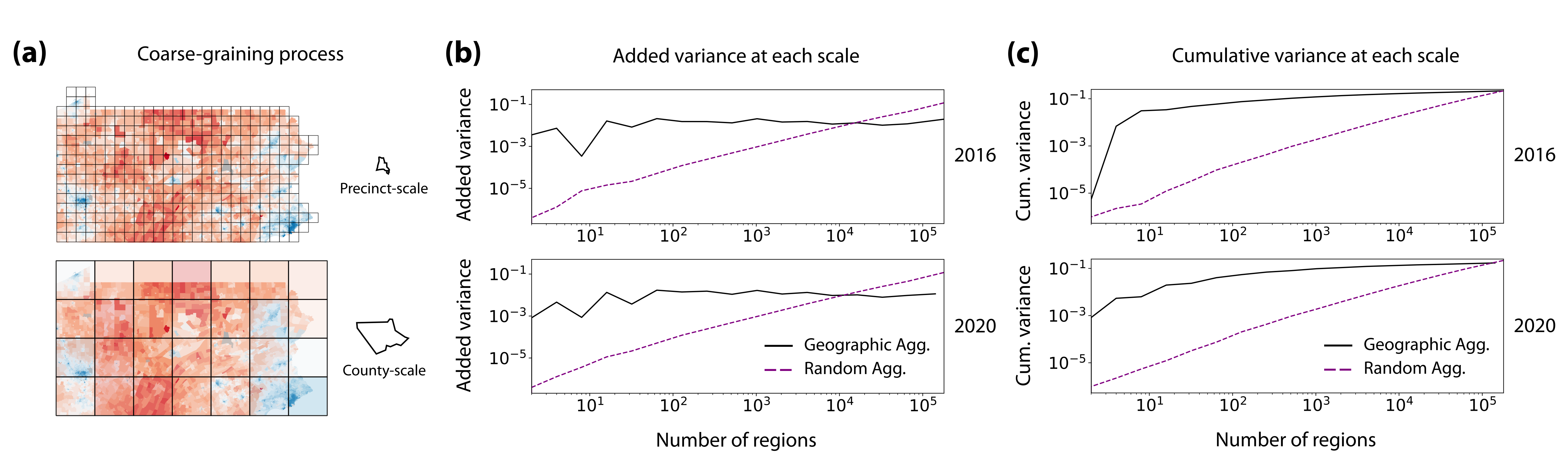}
\caption{\label{fig:precinct} 
(a) Using a map of precinct-level returns~\cite{lai_pennsylvania_2019}, we provide an illustration of a coarse-graining process in which progressively larger numbers of precincts are grouped together.  (b) We employ a variation of this method based on k-d tree partitioning such that each branch contains an equal number of precincts to decompose the opinion variance added at each scale (see equation~\ref{eqn:coarse_grain}) for the 2016 and 2020 presidential elections across the continental U.S. Two lines are shown for each election year: the black line represents the added variance with geographically-based aggregation on precinct-level data \cite{upshot_data2016,upshot_data2020}, while the dashed purple line shows the case when precincts are randomly aggregated with no regard to geography. The logarithmic $x$-axis indicates the number of regions into which precincts are grouped. For example, at the smallest scale (toward the right), no precincts are grouped together and the number of regions is equal to the number of precincts, while at larger scales, precincts are grouped into successively fewer regions. The line of slope 1 in the random aggregation case is characteristic of the central limit theorem. The much smaller slope in the geographic aggregation lines indicates the presence of correlations that persist to large scales. (c) Here, we show the \textit{total} (rather than \textit{added}) variance at each level of resolution for the two elections. As the resolution/number of regions is increased (i.e., scale is decreased/regions are disaggregated), the total variance between regions increases.}

\end{figure*}

Prior works have noted that political polarization is fractal in nature, meaning that it persists at every scale as one zooms into the map \cite{gelman_mathematics_2002,rodden_why_2019}. To quantify this geographic heterogeneity, we begin with a method of breaking up the variance in opinion into the variance arising from each scale---e.g., the variance in opinions among towns, counties, states, and even countries (as is the case of European Parliament elections)---by conditioning the law of total variance upon multiple scales:
\begin{equation}
\label{eqn:coarse_grain}
    \begin{aligned}
        \operatorname{Var}(z)=\mathbb{E}\left(\operatorname{Var}\left(z \mid W_{1}\right)\right)+\mathbb{E}\left(\operatorname{Var}\left(\mathbb{E}\left(z \mid W_{1}\right) \mid W_{2}\right)\right)+  \\ \ldots+\mathbb{E}\left(\operatorname{Var}\left(\mathbb{E}\left(z \mid W_{N-1}\right) \mid W_{N}\right)\right)+\operatorname{Var}\left(\mathbb{E}\left(z \mid W_{N}\right)\right).
    \end{aligned}
\end{equation}
A derivation can be found in appendix~\ref{appendix:total_variance}. Here, $z$ is a random variable that samples across all individual opinions in the country. $W_i$ are random variables that correspond to regions in scale $i$, with $W_N$ corresponding to regions at the largest scale. Each $W_i$ has a probability weight proportional to the population of the region it denotes. As an example, the total variance of the opinion distribution in the U.S., $\operatorname{Var}(z)$, can be broken down into the sum of the average within-city variance of individual opinions, the average within-county variance of the mean opinion of cities, the average within-state variance of the means opinion of counties, and the variance of the mean opinions of U.S. states. This breakdown holds for any random variable $z$, so it can be employed for opinion distributions, election outcomes, etc.

Since political institutions (e.g., city governments, state legislatures, and Congress) operate at different scales, this perspective enables us to quantify the levels at which geographic polarization occurs. As representative bodies capture the total variance at the scale of the election district, one may observe very different levels of polarization in their chambers even if the total opinion variance of the population $\operatorname{Var}(z)$ is held constant. For instance, if next-door neighbors differ greatly but there is little variance among the average political opinion of towns and cities (strong polarization at small scales), local politics may be contentious, with a moderate climate in state and national chambers. Similarly, if towns and cities have divergent opinions but there is little variance between the aggregate opinions of states (strong polarization at large scales), we might expect contentious politics at the state level, with national differences remaining moderate. When there is still substantial variance in opinion when aggregated at the state level, we may expect a tribal Congress and presidential elections that divide rather than unite. 

Figure~\ref{fig:precinct} shows the multiscale breakdown of variance in the U.S. using precinct-level presidential results from 2016 and 2020 \cite{upshot_data2016,upshot_data2020}. This plot can be interpreted as the change in population variance with patch size \cite{smith_empirical_1938, whittle_variation_1956}, illustrating the fractal nature of polarization when one looks at the electoral map with sequentially finer resolution. The distribution of opinions also possesses geographic correlations that persist at large scales: although randomly aggregating regions with no regard to geography will yield central limit-type behavior (as shown by the roughly diagonal lines in (c) where the standard deviation of the average is proportional to the number of precincts $n$ by $1/\sqrt{n}$), the geographically aggregated data has a much smaller slope. This correlation has implications when considering the relative voting power of individuals in smaller and larger states \cite{gelman_standard_2004}.

\begin{figure*}
\includegraphics[width=1\textwidth]{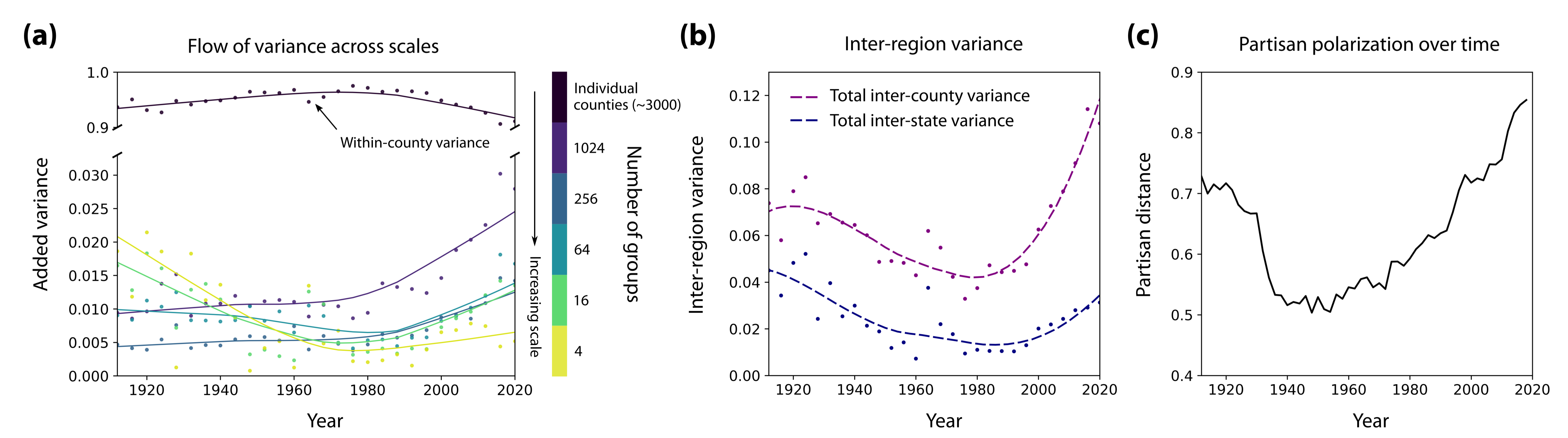}
\caption{\label{fig:variance} (a) Extending the analysis in figure~\ref{fig:precinct}, we use U.S. county-level presidential returns reaching back to 1912 \cite{leip_data} to identify the flow of variance across scales, normalized by the total variance $p(1-p)$ where $p$ is the vote share of the winning presidential candidate. The lines correspond to the added variance at each scale, where the scale is labeled by the number of groups in which counties are geographically aggregated. For instance, the top (dark purple) line corresponds to the variance added by individual counties ($\approx 3000$ groups), and the bottom (light yellow) line corresponds to the variance added at the largest scale, for which the U.S. is divided geographically into four groups. The vast majority of the variance in the system (around 95\%) is contained within the county level. (b) Instead of the variance added at each scale, we can explicitly examine the variance between mean county opinions and the variance between mean state opinions (dashed lines indicate LOESS fits). The total inter-county variance is equal to the sum of all lines in panel (a), excluding the within-county variance. We observe a sharp increase in inter-county variance after 1990, which coincides with the continued increase in partisan distance in Congress as measured by the difference in average DW-NOMINATE scores across the two parties \cite{lewis_voteview_2021} shown in (c).}
\end{figure*}

Figure~\ref{fig:variance} shows that within-county variance has decreased in the U.S. starting around the mid-1980s, translating into a rise in inter-regional variance at larger scales. Although the growth in partisan polarization has its roots in many complex factors, it has coincided with the rise of large-scale variance and what is often described as the nationalization of American politics \cite{hopkins_increasingly_2018}. Splitting the data at the inflection point, the average elector margin in presidential elections (which roughly captures state-scale polarization) from 1916 to 1984 stood at 307, while the average margin from 1988 to 2020 decreased significantly to 138. In the following sections, we continue to explore the implications of this multilevel polarization and discuss the potential relationship between these trends. 


Elections acting at each scale contain only the cumulative variance up to that level, represented by the sum of terms in equation~\ref{eqn:coarse_grain} up to the corresponding scale. Thus, larger-scale elections must resolve more variance. The variance that must be resolved for any electoral or policy decision depends on the scale where that choice is made, independent of any intermediate representation. As we show in appendix~\ref{appendix:total_variance}, in any legislative body, the variance among the legislators \textit{plus} the variance between each legislator and her/his constituents always equals or exceeds the variance of the entire electorate that the legislative body represents.

In other words, for any policy choice at a given scale, the cumulative variance (along the relevant issue dimension) up to that scale must be settled via the electoral system or the agency of public officials. This is a mathematical statement of the arguments advanced by John Milton and James Harrington, who saw the virtues of a federal government in tailoring services to the expressed needs of various subpopulations \cite{beer_notitle_1994}. By customizing policies for each administrative unit, multilevel governance allows variance to be resolved in an efficient way without pushing it up to higher (e.g., national) levels. This is especially true if there is little disagreement within but significant disagreement among administrative units at that geographic scale.

The idea that a fixed quantity of variance must be resolved has implications for how responsibility is distributed in a multilevel system. While elevating the scale of policy implementation may be necessary to match the policy's multiscale complexity with that of the issue it is attempting to address \cite{intro_to_complex_systems2020}, three potential drawbacks should be kept in mind. First, if differences in political opinion arise from genuinely different needs, then the one-size-fits-all approaches necessitated by larger-scale decision-making will be suboptimal. Second, larger-scale policy implementation will require more compromise, with some forced to be bound by the opinions of those residing in completely different areas of the country. Third, attempting to compromise over too much variance in opinion at too large a scale can lead to a destabilizing amount of political polarization.


\section{Elections, social ties, and segregation}
\label{sec:single_dimensional}

Having established a framework to study the geographic distribution of opinions, we now turn our attention to the effects of elections. Here, we introduce a general model of elections to be used throughout the rest of the manuscript: let $S$ be the space of all possible opinions, which we assume can be embedded in a $d$-dimensional opinion space $\mathbb{R}^d$ for some $d$. Then, an election is defined as the process $y: S^n \rightarrow S$ that outputs the opinion of the election winner $y \in S$, where $n$ is the total number of potential voters. Any election, regardless of its structure (number of candidates, voting method, or the presence of a multi-tiered aggregation process like the Electoral College), can be described this way, as a map from a set of citizen opinions to that of an elected official.  In this formulation, candidate positions are endogenous; in other words, the space of possible outcomes of an election is not the discrete set of the positions of candidates who happen to run, but rather the space of all possible candidate positions that could arise.  Thus, the election outcome can vary continuously with the electorate (though it need not necessarily do so---see below), even though any particular election will end up being a choice between a finite number of candidates.  
Defining elections as maps from a given set of electorate opinions $S^n$ requires the opinions be considered at a particular snapshot in time. Thus, there is the implicit assumption that the geographic opinion distribution will be qualitatively similar regardless of the precise time at which the opinions are considered, e.g., one year before the election, one month before the election, or the election day itself. This is a good approximation when voters have relatively coherent and stable opinions in the timescale of interest \cite{ansolabehere_strength_2008}. In this section, we consider phenomena that apply regardless of the dimension $d$ of the opinion space, while in section~\ref{sec:multidimensional} we consider explicitly multidimensional phenomena, i.e., phenomena that cannot be explained if $d=1$. Although the ideas in this section apply for all $d$, we will assume $d=1$ for ease of exposition.

As described in previous work (with slightly different notation) \cite{siegenfeld_bar-yam_2020}, two key failure modes of an election are instability and negative representation. Heuristically, instability refers to the phenomenon in which small changes in the electorate can cause large swings in the election outcome; for instance, the U.S. presidency swung from Obama to Trump, and then Trump to Biden, despite relatively small changes in electorate opinion. These large swings in outcomes correspond to the ``alternate domination of one faction over another" George Washington characterized as a ``frightful despotism" in his farewell address \cite{washington_address_1796}. 

We formalize this notion by defining an election to be \textit{unstable} if the function $y: S^n \rightarrow S$ is discontinuous, i.e., if an arbitrarily small change in electorate opinions can cause a finite shift in election outcome. We can also speak of the magnitude of instability, which corresponds to the magnitude of change in the election outcome that an arbitrarily small change in electorate opinions can produce. Instability can never be directly observed, as it involves a counterfactual in which the electorate has slightly different opinions. Nonetheless, instability can be inferred if swings in outcome from election to election are far larger than could be plausibly expected of swings in electorate opinions.\footnote{U.S. presidential elections from 1944 to 2012 were analyzed and found to undergo a phase transition from stability to instability around 1970~\cite{siegenfeld_bar-yam_2020}.} We should also expect some stochasticity in electorate opinions, which will result in noise of similar or lesser magnitude in the outcomes of stable elections. However, in unstable elections, small fluctuations in electorate opinions (whether treated as random or part of the model) could shift the outcome between radically different candidates. 

Negative representation refers to the phenomenon in which a leftward shift in electorate opinions causes the election outcome to move to the right, or vice versa. For instance, in U.S. elections, a leftward shift in progressive voters may result in their becoming disillusioned with both major-party candidates, causing them to not vote at all or vote for a third-party candidate, which could lead election outcomes to move the right. Another mechanism by which negative representation can arise is through the party primary systems: a shift in voters of one party away from the center may result in a less electable party nominee, which would shift the ultimate election outcome in the opposite direction. Formally, the \textit{representation} of the opinion $x_i$ of an individual $i$ can be defined as the causal effect of a shift in that opinion on the election outcome,\footnote{As shown in reference~\cite{siegenfeld_bar-yam_2020}, this definition of representation generalizes the Owen-Shapley voting power index that is commonly employed in the election literature. There exist specific election functions (i.e., the election outcome as a function of electorate opinions) which recover the deterministic and probabilistic Owen-Shapley indices under their respective assumptions.}

\begin{equation}
    \label{eq:repn}
    r_i = \frac{\partial y}{\partial x_i},
\end{equation}
and can be positive or negative.



We consider instability undesirable for two reasons. First, only a small change in electorate opinions is necessary to significantly change the election outcome, which both makes elections more susceptible to harmful influences (e.g., special interests) and also thereby incentivizes such influences. Second, unstable elections necessarily contain negatively represented opinions,\footnote{For the case of unstable elections, representation may need to be defined for a specific finite change in opinion rather than as a derivative, since the derivative may not exist; see ref.~\cite{siegenfeld_bar-yam_2020} for more details.} to which the election is, perversely, anti-responsive. 
The relationship between negative representation and instability holds regardless of the election mechanism (e.g. the existence of party primaries, the presence or absence of the Electoral College,\footnote
{The fact that the Electoral College and the popular vote can yield such different election outcomes is itself a symptom of electoral instability. A stable election would only be so close so as to be swayed by such factors if the candidates themselves were relatively similar.}
etc.) or the opinion distribution of the electorate.  We direct the reader to reference~\cite{siegenfeld_bar-yam_2020} for more details. 

For ease of notation, we will often consider the election to act on a distribution of electorate opinions $f(x)$, such that the election outcome can be written as $y(f)$ and representation as $r_i=r(f,x_i)$.  This differs from the more general formalism above in that it cannot distinguish between who holds which opinions. 

\subsection{Accounting for social ties}
\label{sec:socialization}

Ideally, democracies are not just mechanisms for opinion aggregation but forums through which citizens and representatives collaborate to reach a common solution. This concept of deliberation is sometimes argued to be the source of democratic legitimacy, embodying the ideas of rational legislation and participatory governance \cite{bohman_deliberative_1997}. Tocqueville described deliberation as being driven by ``enlightened self-interest'' \cite{tocqueville_democracy_2002}: a compulsion for citizens to take into account the opinions of others---particularly those they interact closely with---to maximize long-term payoff. This consideration is more likely to occur among individuals with strong social ties, which in turn are geographically correlated \cite{bailey_social_2018, crandall_inferring_2010}. Scholars have thus argued that deliberation is a scale-dependent phenomena, with Plato and Aristotle famously stating that the ideal size of a polis should not exceed 5040 citizens \cite{plato_plato_2016}. Indeed, a key argument made for federal governments is that they combine the ability for small states to foster participation with the advantages of a large republic \cite{inman_democratic_2020}.

Here, we develop a general model to explore how polarization and representation are affected by social ties. Unlike previous studies that examine how social networks affect information transfer \cite{stewart_information_2019} and opinion formation \cite{holley1975ergodic, hegselmann2002opinion, borghesi_spatial_2010}, we do not make assumptions about how preferences diffuse and evolve. Rather, we take the opinions of voters as given (as described in section~\ref{sec:single_dimensional} above) and impose on them a change in voting behavior based on the set of social neighbors to capture the multiscale effects of these interactions. We show that although social ties can be beneficial in encouraging deliberation, this type of interaction may conversely aggravate polarization if insular patterns of political socialization emerge \cite{johnston_question_1991}. In section~\ref{sec:seg}, we use this model to understand the geographic interactions between social ties and elections held at various levels.

A citizen's \textit{effective opinion} $x'$ is defined as a weighted average of the opinions of themselves and their neighbors:
\begin{equation}
\label{eq:social_opinion}
    x'_i = \sum_{j} T_{ij} x_j,
\end{equation}
where $T_{ij}$ is some social connectivity matrix defined such that for each individual $i$, $\sum_j T_{ij}=1$ (so that translational invariance is maintained), yielding an effective opinion distribution $\hat{f}(x')$ on which the election acts.\footnote{More generally, $T_{ij}$ could also be negative.  Negative weights capture the effect that socializing with certain people causes one to vote further away from rather than closer to their ideal points \cite{macinnis_how_2015}. Negative and positive weights correspond to the effects of threat and contact theories of interpersonal interactions respectively. These notions describe simultaneous and opposing forces but often operate on different spatial scales; contact requires frequent inter-personal interactions while threat may be perceived on a large scale because economic or political competition may operate at a state or national level \cite{biggs_explaining_2012}. Similarly, if social media---which connect people on a national scale---do lead to a more negative evaluation of differing opinions \cite{bail_breaking_2021}, we may consider a model in which larger-scale connections have negative weights while smaller-scale, face-to-face interactions have positive weights, strengthening the effects described in this section.} In other words, in the presence of social ties, the election outcome is given by $y(\hat{f})$ rather than $y(f)$. Because the model does not make explicit assumptions about opinion dynamics and can accommodate a wide variety of social network structures (encoded by $T_{ij}$), it can be expected to be applicable to a wide variety of real-world scenarios. 

We expect social ties to make representation more equitable. Indeed, representation can be calculated to be:
\begin{equation}
    r_i=\frac{\partial y(\hat{f})}{\partial x_i}\bigg\vert_{x_{k \neq i}} =\sum_j T_{ji}\frac{\partial y(\hat{f})}{\partial x'_j}\bigg\vert_{x'_{k \neq j}} =\sum_j T_{ji}r(\hat{f}, x'_j).
\end{equation}
When social connectivity $T_{ij}$ is positive $\forall$ $i, j$ (as it is in an deliberative democracy), representation will tend to be more evenly distributed. However, in general, the total representation will not increase, and individual representation will remain $\mathcal{O}(1/N)$, where $N$ is the size of the electorate. For instance, for differentiable elections, we have the exact result $\int^{\infty}_{-\infty}f(x)r(f,x)dx = 1$~\cite{siegenfeld_bar-yam_2020}.

We can also define \textit{social representation} as the change in election outcome with respect to the effective opinion, holding the selfish opinions of everyone else constant:
\begin{equation}
    \frac{\partial y}{\partial x'_i}\bigg\vert_{x_{k \neq i}} =
    \frac{\partial x_i}{\partial x'_i} \bigg\vert_{x_{k \neq i}} \frac{\partial y}{\partial x_i}\bigg\vert_{x_{l \neq i}}  = \frac{1}{T_{ii}}r_i.
\end{equation}
This social representation measure captures the positive-sum nature of individuals taking into account each others' preferences. In a hypothetical group in which everyone valued the opinions of others equally, every individual would have a representation of 1, capturing the fact that if the preferences of all citizens were equally weighed by all individuals (such that everyone's effective preferences were the same), the government would be fully responsive to that effective preference.    

\subsection{Multiscale effects of geographic segregation}
\label{sec:seg}

\begin{figure*}
\includegraphics[width=1\textwidth]{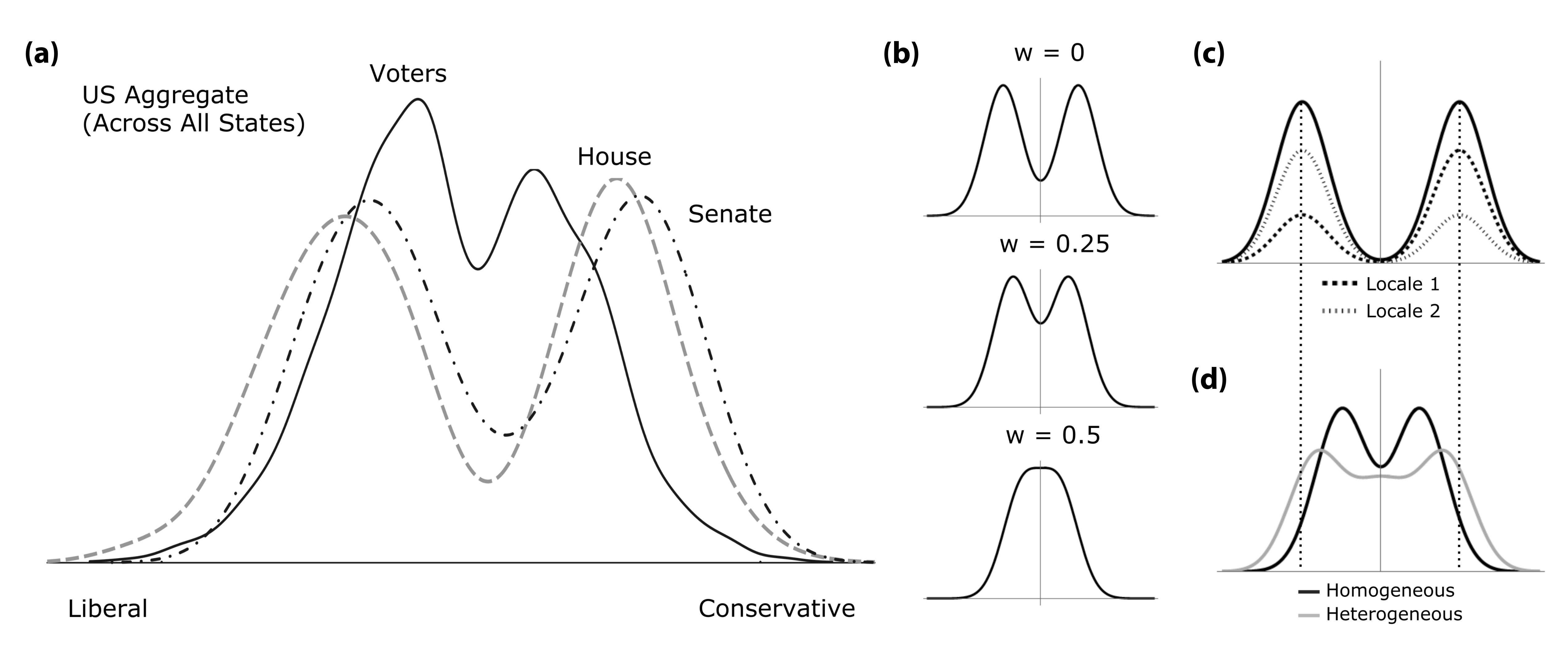}
\caption{\label{fig:socialization} (a) We illustrate the effects of social ties using a bimodal opinion distribution, similar to the aggregate voter and representative ideal points across the U.S. as estimated by Bafumi and Herron \cite{preference_aggregation, gelman_red_2010}. (b) Within a single locale, social ties among all the members reduce the variance of each component distribution and shift their means closer to each other. For sufficiently strong social ties (parameterized by $w$), the election becomes unimodal, as seen when $w = 0.5$. (c) Next, we extend this framework to the multiscale case with two locales. The two locales may be completely identical, or they may each be politically segregated such that one is biased toward the first peak (black dashed line) and the other toward the second peak (gray dashed line), e.g., one being a majority-Democratic and the other a majority-Republican jurisdiction. The overall opinion distribution (and thus the total variance) is the same in both cases. (d) However, as a result of local social ties, the segregated (heterogeneous) system---in which much of the variance is between the locales---will display more polarization than the homogeneous system---in which all of the variance is within the locales.}
\end{figure*}

As social ties distribute the representation of individuals, they have the potential to reduce the amount of negative representation. They can also potentially decrease the degree to which the election is prone to instability.  Whether these benefits are realized, however, depends on the way in which social ties are distributed across the opinion distribution, which will in turn depend on the geographic distribution of both social ties and political opinion. 

Social ties that span an electorate reduce the effective variance of its opinion distribution. Consider the opinion distribution in figure~\ref{fig:socialization}b consisting of two normally-distributed subpopulations. If we consider a connectivity matrix $T_{ij}=w/(n-1)$ for $i\neq j$ where $n$ is the size of the electorate (and thus $T_{ij}=1-w$ for $i=j$), the opinion distribution is transformed, reducing the distance between the means of the two subpopulations by a factor of $1-w$, while also decreasing their respective scales by the same amount. More generally, homogeneous social ties decrease the effective variance regardless of the precise form of the opinion distribution; see appendix~\ref{appendix:network_structure}. Thus, to the extent that social ties are geographically localized, this analysis implies that electoral polarization is less problematic locally than at larger scales. 

However, while social ties decrease instability within well-connected locales, they can also increase the overall polarization of the system depending on the structure of social connections. Political homophily---in which citizens associate themselves with people of similar political views---has been observed to be a key process in many social networks \cite{huber_political_2017}. Consider two groups of separate ideologies that are socially disconnected from one other. This is an extreme form of affective polarization, wherein citizens become unwilling to socialize across party lines due to the emergence of partisanship as a social identity \cite{iyengar_origin_2019, finkel_political_2020, druckman_affective_2021}. Social ties cause the effective opinions within each group to cluster more sharply around their respective means. For example, calculating the effective opinion distributions using a bimodal distribution, the centers of the two Gaussians stay the same, but since their scales are decreased to $\hat{\sigma} = \sigma (1-w)$, the effect of social ties tends to increase instability (appendix~\ref{appendix:network_structure}). Growing instability and the structure of social ties can be self-reinforcing across the span of several elections, especially when institutions like strong party systems begin to steer public opinion \cite{chapel_agenda_setting, cox_setting_2005}.

Thus far, we have discussed the effects of social ties in a hypothetical election where voters are either well-connected or fully segregated according to partisan identity. We can generalize this model across multiple levels---following the framework introduced in section~\ref{sec:coarse-graining}---by assuming that social ties are present with varying strengths at each scale. This assumption is based on two mechanisms. First, elections at each scale (e.g., mayoral or gubernatorial elections) mediates interactions between voters. Second, despite the complexity of human networks, social interactions---particularly high-salience ones that involve face-to-face exchanges---are often geographically correlated. Therefore, we introduce a hierarchy of coupling weights $w_i$, each corresponding to the density of social interactions among citizens within the same scale-$i$ region. 

This model yields two key results. First, since the degree of social ties varies at different scales, the effective total variance in a country is affected not only by the opinion variance of the electorate, but how those differences are distributed across levels. Specifically, the terms in equation~\ref{eqn:coarse_grain} are transformed as
\begin{equation}
\label{eqn:xformedvar}
\begin{aligned}
    \operatorname{Var}(z')&=\mathbb{E}\left(\operatorname{Var}\left(z \mid W_{1}\right)\right) (1-\Sigma^{N+1}_{i=1}w_i)^2 \\ &+ \mathbb{E}\left(\operatorname{Var}\left(\mathbb{E}\left(z \mid W_{1}\right) \mid W_{2}\right)\right)(1-\Sigma^{N+1}_{i=2}w_i)^2 \\ &+ \dots +  \operatorname{Var}\left(\mathbb{E}\left(z \mid W_{N}\right)\right) (1-w_{N+1})^2,
\end{aligned}
\end{equation}
where $z'$ is a random variable sampling over the effective opinions and $W_i$ are once again random variables corresponding to regions at scale $i$. (Note that the scale $N+1$ corresponds to the entire country, and thus $w_{N+1}$ denotes the strength of nationwide social ties.) If social ties are stronger at smaller scales, it would be preferable for a larger portion of the total variance to be present in those levels (see figure~\ref{fig:socialization}d).

Second, we find that increasing segregation at any particular scale, such that opinions in regions below that scale are made more homogeneous and opinions above it are made more heterogeneous, can stabilize local elections while destabilizing larger-scale ones. This type of segregation could occur through partisan sorting, urbanization, or reverse effects \cite{bishop_big_2009, martin_does_2020, kaplan_partisan_2020}. Furthermore, increasing the relative strength of local ties in this situation is at the detriment of larger-scale elections, essentially aggravating the effect of insular communities. This result is a multilevel generalization of our previous finding that segregating social ties across party lines increases the likelihood of electoral instability. A detailed discussion can be found in appendix~\ref{appendix:geograpical_segregation}.

Although this model of social ties is certainly a simplified one, it captures the idea of deliberation that is central to many arguments for federalism and participatory democracy. If people of different political opinions were geographically randomly distributed, election instability would be significantly reduced, since any individual---even if they are upset that their own views were not represented in government---would be surrounded by many people whose views are represented and would recognize the need for compromise. In this limit, political opinions would be well-described by a mean-field theory (i.e., any individual opinion could be described as the mean opinion plus some uncorrelated noise) and so could be well-represented by a single instrument (e.g., the national government), although other considerations may still favor more local forms of representation and policy-making.  The inability for a top-heavy political system to stably represent the U.S. electorate can be viewed as a consequence of  a geographic opinion distribution that is in reality not so well-mixed.\footnote{Without deliberation to cut down political polarization, a Congress consisting of multiple members does little to ameliorate this problem, as Congress must still come to a single decision for the entire nation on any given piece of legislation. As discussed in section~\ref{sec:coarse-graining}, the problem of compromise is simply shifted from the electorate to Congress.} 


\section{Multidimensional preferences}
\label{sec:multidimensional}

\begin{figure}
\includegraphics[width=0.5\textwidth]{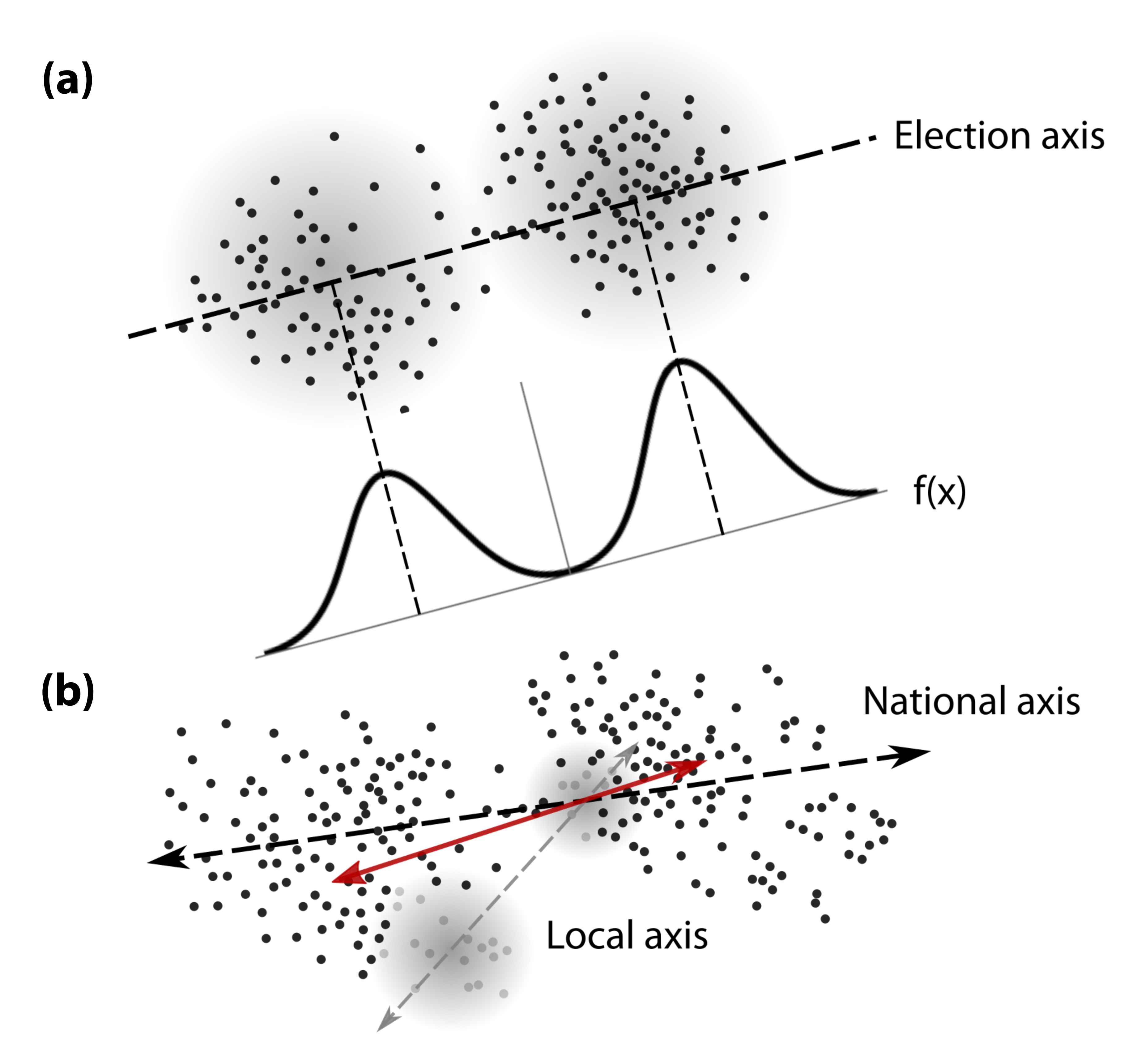}
\caption{\label{fig:axes} (a) As each candidate aims to build a coalition of voters, the contest occurs on an axis that approximately spans the direction of maximum variance. The election can be approximated as acting on the projection $f(x)$ of the multidimensional opinion distribution onto this axis. (b) The same process can be repeated for a subset of the whole electorate (marked in gray), which corresponds to a local election. The introduction of interactions between the two axes may result in a local election axis (marked in solid red) that differs from the axis that would maximize the projected variance of the local opinion distribution.}
\end{figure}

Up to this point, we have discussed behaviors whose essential forms can be described in a single-dimensional opinion space. While this is a reasonable simplification for isolated elections---where our results on the multiscale composition of variance and the effect of social ties hold regardless of dimensionality---a multidimensional space is needed to model how different elections interact. We now explore phenomena that cannot be explained without explicit reference to a multi-issue space, beginning with the problem of how opinions that may lie in a high-dimensional space can be measured.  

\subsection{Election axes}
\label{subsec:election_axes}

Measurements of political opinions are projections of voter preferences along the directions spanned by the set of instruments (e.g., poll questions) used. This leads to two immediate observations. First, unless a set of basis vectors spanning the whole space is constructed, the measurement does not yield complete information on the opinion distribution. Second, it is generally difficult to conclude that there is no mass polarization as a polarized distribution may not have a high variance (or not appear as bimodal) when projected onto a smaller subspace. Even a comprehensive study that integrates polling data on a large number of opinion dimensions will not necessarily uncover the full structure of the space.

Care needs to be taken when operationalizing the opinion distribution with poll-based measurements. Besides being inadequate for fully reconstructing the opinion space, polls are typically constrained along a number of natural axes: easily pollable issues or those often discussed by political elites. This idea has been explored empirically. For instance, Broockman (2016) showed that commonly used ideological scores are poor measures of policy preferences. Analysis using a wider range of measurement axes finds legislators to be ``similarly moderate as voters, not more extreme" \cite{broockman_approaches_2016}. This result aligns with the finding by Ansolabehere et al.\ that apparently unstable and incoherent voter opinions are manifestations of measurement error. Increasing the number of survey items improves the stability of opinion, steadily approaching that of party identification \cite{ansolabehere_strength_2008}.

\begin{figure}[b]
\includegraphics[width=0.47\textwidth]{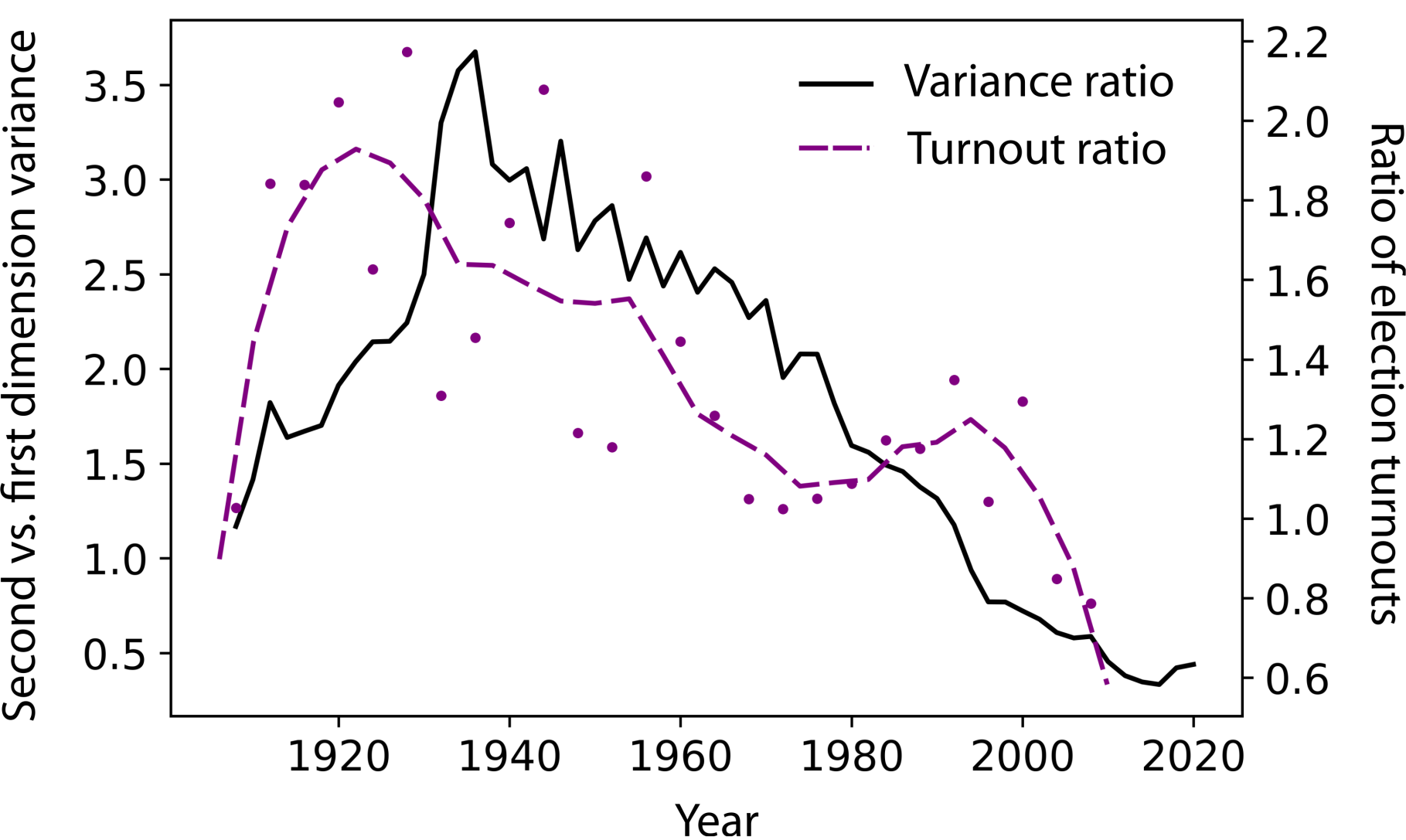}
\caption{\label{fig:salience} Relative salience of local and national elections and the dimensionality of congressional voting patterns. The former is measured via the ratio of voter turnout of American gubernatorial to presidential primaries held on the same year (dashed line indicates the LOESS fit).  The latter is measured via the variance in DW-NOMINATE scores on the second axis divided by the first axis, as computed from congressional roll-call data \cite{lewis_voteview_2021}; higher values of this variance ratio indicate a greater importance of issues outside the liberal-conservative axis.}
\end{figure}

The idea of a measurement axis can be generalized by modeling elections as measurements along the ideological positions of the candidates, yielding an \textit{election subspace} spanned by the candidates $\mathbb{R}^{\operatorname{min}[c-1, d]} \in \mathbb{R}^{d}$, where $c$ is the number of candidates in a given election. In contrast to poll questions, candidates can more flexibly take positions across a broad range of issue dimensions, many of which may be illegible. 
Our analysis thus far applies to multilevel democracies generally and does not depend on the nature of the electoral system. For simplicity, we now restrict ourselves to two-party elections that operate on an \textit{election axis} defined by the line containing the positions of the two candidate (though, as we will discuss, such an axis with $d = 1$ is still a useful concept in multi-candidate elections). Despite the complexities of elections in such a high-dimensional space, two-party elections can be summarized by the the one-dimensional axis $\hat{\mathbf{e}}$ spanned by the two candidates, together with the one-dimensional opinion distribution $f(x)$ created from the projections of every opinion position onto $\hat{\mathbf{e}}$, as illustrated in figure~\ref{fig:salience}. We might expect this election axis to often roughly coincide with the axis of political discourse occurs, with $f(x)$ representing the distribution of opinions concerning this discourse. 

Defining an election subspace does not impose any additional assumptions onto a multidimensional election. In particular, the election axis represents a choice of how to abstract the election process rather than an assumption about the election mechanism. Polarization can then be assessed from the one-dimensional opinion distribution that arises from the projection of electorate opinions onto the line spanned by candidate positions. This coarse-graining enables us to capture the relevant large-scale behavior of the election---regardless of the details of how outcomes may emerge from an multidimensional space---and build up to a picture of cross-level interactions in section~\ref{subsec:axis_socialization}. 

We now examine one particular model, in which the election is produced by dividing the opinion space across the dominant cleavage line with one party residing in each cluster. A simple way to capture this process mathematically is via $k$-means clustering. When $k=2$, two natural clusters are formed by minimizing the mean-squared distances between the cluster centroids and their surrounding samples. We can interpret this as a process in which candidates attempt to minimize the overall ideological distance between them and their supporters. Such a process sets a general axis of discourse along the means of the two clusters $\vec{\mathbf{\mu}}_1 - \vec{\mathbf{\mu}}_2$, the normalized version of which define as the election axis $\hat{\mathbf{e}}$. This roughly corresponds to splitting the electorate along the axis of greatest variance in the case of a two-party election; see appendix~\ref{appendix:candidate_selection}.

More generally, the election axis lies perpendicular to the traditional notion of a partisan cleavage \cite{lipset_cleavage_1967, miller_activists_2003}.\footnote{In an Euclidean space, dividing voters according to which party they are closest to results in a boundary that is  perpendicular to the election axis. When there are multiple parties, these cleavages partition the opinion space into Voronoi cells, each containing the set of voters closest to the corresponding centroid.} If the partisan cleavage is aligned with the \textit{dominant} social cleavage, the election takes place along the axis of highest variance. This is true especially when parties start to shape the dominant social cleavage (e.g., via homophily, as discussed in section~\ref{sec:socialization}) over the course of multiple elections, unless the parties or issues are going through significant reorientation \cite{miller_activists_2003}. Consequently, the polarization measured along the electoral axis will be greater than or equal to those found along an arbitrary set of measurement axes. This provides a plausible explanation for why surveys of both partisan and elite polarization are often trailed by measurements of mass polarization: candidates are simply positioned along the axis with the greatest projected variance. The social cleavage definition also makes the election axis a useful construct in multi-candidate elections as it indicates the primary direction of discourse.

\subsection{Interactions among election axes at different scales}
\label{subsec:axis_socialization}

Next, we explore how multilevel interactions affect political polarization using the concept of election axes described in the previous section. Within a democratic system, elections are not independent from each other because municipal, state, and national politics are nested \cite{golder_multi-level_2017}. Institutional effects pull lower salience elections toward the direction of discourse in more dominant contests \cite{reif_nine_1980, coattail}. Interactions may also be driven by the existence of a strong party system \cite{cox_legislative_1993}, shared funding resources, the delocalization of news media \cite{hopkins_increasingly_2018}, or the simple fact that the same politicians are often active across multiple scales of government. Such effects are more pronounced in countries with advanced party systems but have been described across a wide range of democracies \cite{jones_nationalization_2003}. 

Rather than posit specific models, we use the election axis as a coarse-grained description of multilevel dynamics through which the effects of various interactions are included. The first way contests can affect each other is via issue activation: elections amplify the difference in candidate opinions as campaign messaging and media coverage focus on how their positions diverge. For candidates positioned at $\vec{x}_1$ and $\vec{x}_2$, communications are typically aligned with their differences, described by the unit vector $\hat{e} = (\vec{x}_1 - \vec{x}_2)/|\vec{x}_1 - \vec{x}_2|$. In a system with multiple elections, a highly salient contest (such as a presidential election) can activate issues along $\hat{e}$ and focus public attention onto that axis, even in mayoral or gubernatorial elections where it might not represent the primary issues relevant to that level of governance. This effect persists even in elections where all the candidates belong to the same party.

Second, partisan and institutional effects can constrain candidate positions. We employ the formalism developed in section~\ref{sec:socialization} to model these forces as ties among members of each political faction across scales (such as ties between local and national candidates of the same party). Consider two elections which, without mutual interactions, span election axes $\hat{e}_a$ and $\hat{e}_b$. As detailed in appendix~\ref{appendix:axis_interaction}, within-party ties across the two elections shift the axes toward each other, bringing them to $\hat{e}'_a = w_a \hat{e}_a + (1-w_a) \hat{e}_b$ and $\hat{e}'_b = w_b \hat{e}_b + (1-w_b) \hat{e}_a$ respectively, where $w_a, w_b \in [0, 1]$. The effect of these ties is that the angle subtended by the two axes is now reduced. Generally, $w_a \neq w_b$, as interactions between the two elections may be asymmetrical, especially if they occur at different scales, or if one is more politically salient than the other. This asymmetry can also be driven by the relative distribution of policy responsibilities between the two elections.

Such interactions consist of what Hijino and Ishima (2021) termed ``multi-level muddling," wherein candidates adopt messages that appeal to performances and issues in levels of government other than the one in which they are seeking office \cite{hijino_multi-level_2021}. In the U.S., for instance, state-level candidates often focus on issues that resonate across the country \cite{carr2016origins}, sometimes even de-emphasizing policy issues that are under their purview \cite{grumbach2018backwaters}.

Ultimately, both media framing and institutional ties have the same effect: the direction of discourse becomes aligned across elections at different levels, reducing the angular dispersion between their axes. Although interactions in a multilevel system can be described more generally (as elaborated in appendix~\ref{appendix:multilevel_axis}), we shall demonstrate these effects using on a simple system with local-national ties. For instance, suppose a country has a set of pre-interaction local axes $\hat{\textbf{e}}_{\ell} = (\hat{e}_{\ell, 1}, \hat{e}_{\ell, 2}, \hat{e}_{\ell, 3}, \dots)$---corresponding to congressional elections---and a national axis $\hat{e}_N$, corresponding to the presidential election.\footnote{We note that although congressional elections elect a candidate for federal office, we refer to their election axes as local since such axes coarse-grain regional electorates.} The existence of cross-level interactions has two key implications.

First, these interactions mean that congressional representatives---each chosen from different regional electorates---will be drawn from elections with more aligned axes if multilevel interactions are strong and national salience is high. If the national axis $\hat{e}_N$ has a strong pull, winning candidates are likely to be clustered along the national axis of discourse, rather than being scattered across the ideological space due to the diversity of local and regional concerns. The way in which this clustering leads to congressional polarization will be discussed in more detail in section~\ref{sec:madison}. Furthermore, to the extent that state politics are influenced by these national concerns, state governments will cease to serve as a check against national polarization (for instance, state legislators often draw voting districts and pass voting laws in line with their national party). 

Second, as outlined in the previous section, the election axis determines how the multidimensional opinion space is projected onto the election as a one-dimensional distribution $f(x)$. A tradeoff can arise in the variance projected against national and local elections depending on the relative salience between the two. 

If the salience of local politics is high, then each local election freely picks its direction of discourse $\hat{e}_{\ell, i}$, while the national election---with a weaker agenda-setting capacity---is pulled by an aggregate of these local axes. The pull results in an effective axis $\hat{e}_{N}'$ that may not correspond exactly to the national cleavage. In this case, local elections tend to maximize their projected variance, while national elections do not.

On the contrary, if national salience is high, presidential elections would occur against the main national cleavage, setting up a contest along the direction of maximum variance. Because the dominance of national discourse pulls local candidates away from strictly local contests, opinion variance against their axes can be lower than what would otherwise be observed against the optimal local divide. In such a climate, one may observe increasing polarization at the national level while local elections simultaneously become more dominated by single parties \cite{schleicher2008}. 

As Gelman (2014) discussed, the recent appearance of close elections is relatively unusual in American history. For instance, in less nationalized periods like the early twentieth century, Democrats were largely content with controlling the urban political machines and the American south \cite{gelman_twentieth-century_2014}. Parties resorted to national politics only as a brokering mechanism. The rise of federal spending eventually incentivized them to invest in national contests whenever possible \cite{ferguson_golden_1995}, leading to an increase in the number of close contests. The recent multilevel effects of nationalization may be seen in party platforms: using automated and manual content analysis, Hopkins et al. (2022) found that within-party variation among local platforms in the U.S. has decreased significantly since the mid-1990s, while between-party differences in the topics discussed diverged over the same period \cite{hopkins_many_2020}. 

Although these two cases highlight a tradeoff between national and local polarization, the scale at which highly divided contests occur can matter in the long run. In particular, polarization at larger scales can have ripple effects not present at smaller scales by deepening the national cleavage across multiple elections. Members of each party may further congregate in opinion through internal interactions and agenda setting \cite{cox_setting_2005}, forming a positive feedback loop that aggravates the social divide.

We can further describe the effect of multilevel constraints on democratic accountability using the concept of representation from section~\ref{sec:single_dimensional}. Assuming the derivative exists, we can generalize equation~\ref{eq:repn} to a multidimensional space, writing the representation of opinion $x_i$ as $r^i_{\mu\nu} = \partial y_\mu/\partial (x_i)_\nu$ \cite{siegenfeld_bar-yam_2020}. This is a rank-two tensor where the first index corresponds to the direction of change in the election outcome and the second index corresponds to the direction of opinion change. 

Any change in opinion can be broken down into its component along and orthogonal to the election axis $\hat{e}$, enabling us to split $r^i_{\mu\nu}$ into on-axis and off-axis representation; see appendix~\ref{appendix:multidimensional_representation} for details.

For instance, if representation occurs only along the election axis, perhaps due to sparse public discourse along orthogonal directions, a regional election can offer representation on local issues only if the election axis is aligned with such matters. The nationalization of regional elections may prevent such an alignment~\cite{hopkins_increasingly_2018}. Such a behavior is not predetermined by our model, however. There may also be other cases in which candidates are free and willing to move in orthogonal directions in response to changes in public opinion, offering an avenue for representation even if the on-axis contest is unstable and, by extension, contains negatively represented opinions \cite{siegenfeld_bar-yam_2020}.

\subsection{Guarding against polarization}
\label{sec:madison}

In this section, we consider in more detail the effect of multilevel interactions on legislative bodies. As mentioned in the previous section, as the salience of national elections increases, the election axes of regional elections become focused in the same direction and politicians congregate around their respective clusters. For example, in the U.S., higher salience in national politics will lead to a more one-dimensional Congress in a two-party system. Conversely, greater salience in local politics---and thus greater freedom in selecting ideal platforms---results in a more varied Congress, with variance distributed along different dimensions.

The framers of the American constitution expected state-level loyalties to far outweigh those to the new nation so that local attachments could counterbalance the centralizing tendencies of a large republic \cite{levy_federalism_2007}. Madison wrote about this in Federalist No.\ 10: in a world where local opinions are distributed among different issue dimensions, factions are suppressed ``by their number and local situation," leading them to ``lose their efficacy in proportion to the number combined together." This variation, he argues, prevents the formation of a dominant faction at the national level. As the premise that local attachments trump national ones breaks down, we have observed an erosion of local authority and the nationalization of regional politics \cite{hopkins_increasingly_2018}. The purported ability of a federal system to insulate a country from factionalization, therefore, has also fallen short.

We can formulate the effects of multiple issue dimensions more precisely with a simple mathematical model. Consider an opinion space where all voters have equal extremity such that they are equidistant to the mean. If variation in opinions only exists across one issue dimension (such as the case of extreme nationalization), there are exactly two groups, each located a distance $r$ from the center. In this situation, the total variance in the distribution is simply $\sigma^2 = \sum_i (x_i-\bar{x})^2/P = r^2$, with $P$ being the total population.

Next, we add issue dimensions to the system. If there is an equal amount of variance on each axis---corresponding to a case where disagreement occurs across a multitude of issue dimensions---then opinions are distributed evenly on a spherical surface of radius $r$ and dimension $n-1$ where $n$ is the number of issue dimensions. Since the average squared distances between the opinions and the mean is $r^2$, this yields a covariance matrix of the form $K_{ab}= \sigma^2 \delta_{ab}$ with $\sigma^2 = r^2/n$. This implies that in a system where all opinions are equally far from the center, the variance $\sigma^2$ projected onto any one-dimensional election axis decreases as the number of directions in which opinions vary increases.\footnote{By a similar argument, Rodden (2021) showed that affective polarization intensifies when more issue dimensions are added \cite{rosenbluth_keeping_2021}. While this may at first seem to be in conflict with our result, these two conclusions represent two sides of the same coin: we hold overall polarization constant (by placing everyone on a sphere of equal extremity to the mean), while Rodden's model assumes that partisan polarization (i.e., the variance projected onto the 1D axis between the parties) is held constant.} Thus, the effectiveness of a pluralistic system in guarding against polarization is reduced if opinions collapse to a low-dimensional space. As the salience of national elections becomes high, the opinions of representatives in Congress become more one-dimensional, meaning that the effects of polarization are significantly more pronounced. A multilevel democracy would be more effective in guarding against factions if local concerns were more salient than national ones, as Madison seemed to have assumed.  

The dimensionality of Congress can be indirectly measured through the explained variance of the first and second principle axis in roll-call votes. Although partisan alignment is likely driven by a range of complex issues, figure~\ref{fig:salience} shows that dimensional collapse has occurred during the same period as the relative salience of regional politics---as measured by the ratio of turnout in gubernatorial to presidential elections---has decreased.

The domination of national issues over local ones can diminish the diversity of opinions, intensifying polarization along a single direction. It also suggests a vicious cycle: the more salient national issues are, the more attention is paid to national governance, and the more citizens expect issues to be solved by the national government. The national government may then take on more responsibility relative to local ones, which in turn results in greater national salience. 

\subsection{Opinion aggregation across multiple issue dimensions}
\label{subsec:multidimensional_aggregation}

The multilevel breakdown of variance presented in section~\ref{sec:coarse-graining} is directly applicable to individual issues when considering a multidimensional space of opinions. Just like the single-dimensional case, each issue can have a varying degree of cumulative variance at each geographic scale. A country can have disagreements on economic policy at the smallest scales (e.g., between individuals and their neighbors) while opinions on gun ownership are locally homogeneous and significantly divided at the largest scales (e.g., between north and south).

The set of cumulative variances of these issues determines the natural axis of discourse for an election at a particular scale (before interactions with other election axes are considered). Under certain assumptions (see appendix~\ref{appendix:candidate_selection}), if the distribution of opinions within an electorate at scale $i$ is summarized by a covariance matrix $\tilde{K}^i$, the election axis is approximated by the top eigenvector of this matrix, with the associated eigenvalue corresponding to the opinion variance projected onto that axis. The multiscale breakdown of variances across each issue dimension for a country (i.e., the multidimensional generalization of figure~\ref{fig:precinct}) can be written as the sum of the added covariance matrices at each scale.

We illustrate the effect that issue aggregation can have on a country with two cases. First, consider a situation where differences in electoral opinion are concentrated along similar axes for regions above scale $i$ (for example, the added variance above the county scale is mainly along the standard liberal-conservative axis), whereas the added variance below scale $i$ is dispersed along a wide range of issue dimensions (e.g., people in each town differ greatly in their preferences for education and policing). In such a country, local elections would not have a dominant axis of discourse, but a clear issue dimension with high variance emerges nationally. The flip side can also occur: if disagreement among citizens occurs locally along a small subset of issue dimensions but exists nationally on a wide range of issues, local elections will be more polarized with stable election axes but larger-scale elections will lack a dominant axis. This latter case reflects the phenomena described in section~\ref{sec:madison}, wherein the diversity of issues relevant at the national scale guards against polarization. 

While the implementation of any given policy may be easiest at a particular level of governance, one must also take into account the multiscale distribution of opinions to ensure that polarizing issues, in Madison's words, ``lose their efficacy." Moving issues that are most contentious at a particular electoral scale to other levels of government can serve as a counterweight against the emergence of a highly polarized dominant election axis.

\section{Summary}

It has been argued that federal democracies offer advantages in encouraging political participation, the protection of individual rights and liberties, and economic efficiency \cite{inman_democratic_2020}. However, these are not intrinsic properties of federal systems. Realizing these benefits depends on the relationship between different levels of governance, the multiscale complexity of the policy problems each has to tackle, and the geographic structure of the electorate. While the link between federal governments and their policy environments has been extensively studied, our analysis provides a new framework for describing how the third pillar---the multilevel distribution of political opinions---couples with democratic processes.

As we outline in section~\ref{sec:coarse-graining}, varying degrees of polarization occur along different issue dimensions at each level of governance. This leads to a tradeoff in assigning policy scopes by implementational efficiency alone: in addition to considering the advantages and disadvantages of enacting policy at a given level of governance, one must also take into account the amount of polarization that needs to be resolved at that level. Factoring in the idea that deliberation and political participation are only optimal at certain scales---as scholars from Plato to Tocqueville have argued---we show how geographic segregation affects political polarization and representation, and how the distribution of variances at different scales changes the stability of local and national elections. We then explore the structural factors of national-local relationships by extending the framework of multilevel polarization to a multidimensional opinion space. We demonstrate that increasing national salience can result in elections occurring predominantly along a single one-dimensional axis, spoiling the purported insulation against polarization that a federal system provides.

These multilevel considerations suggest a strong link between political nationalization and polarization. Both voter turnout and engagement in local politics have decreased significantly over the past few decades \cite{oliver_local_2012, einstein_pushing_2016, schaffner_hometown_2020}, while polarization at larger scales (e.g., the variance between counties, the variance between congressional districts, and the variance between states) has grown. Resolving more issues via local elections can help reduce instability and increase representation by distributing polarization more evenly across scales and leveraging social effects to encourage deliberation. Doing so can also transfer some of the political salience of national elections to state and local ones and increase local turnout.

This paper has touched on a wide range of topics, with the purpose of providing new mathematical and conceptual frameworks for future research, rather than definitive answers. Our unifying theme is that in addition to matching the comparative advantages of each level of government with its policy environment, it is also essential to consider how differences in opinions are distributed geographically. When political preferences are geographically clustered, devolving the relevant policies to lower geographic levels can reduce the risk of polarized and unstable national elections. Only with a careful balancing of the policy issues tackled at each level of government can the full advantages of a federal system be realized. 

\begin{acknowledgments}
    We thank Johnathan Rodden for conversations on fiscal federalism, polarization, and multidimensional opinion spaces, and Shigeo Hirano for insight on the nationalization of multilevel politics. We would like to express our appreciation to Deb Roy for his comments and support of this project at the Center for Constructive Communication. The authors also gratefully acknowledge the U.S. Office of Naval Research, the National Science Foundation Graduate Research Fellowship Program under grant no. 1122374, and the Hertz Foundation for partial support of this research.
\end{acknowledgments}

\bibliography{references}


\appendix

\section{Multiscale total variance}
\label{appendix:total_variance}

As described in section~\ref{sec:coarse-graining}, we can break up the total variance of a random variable $z$ into the variance arising at each geographic scale. In the context of multilevel polarization, $z$ may correspond to political opinion of a random voter within a country.  The country can be partitioned into a nested hierarchy of regions of increasing scale; as just one example, scale 1 could correspond to precincts, scale 2 to counties, and scale 3 to states.  Letting $W_n$ be a random variable denoting regions at scale $n$ with probabilities proportional to their populations, the law of total variance yields     
\begin{equation}
\label{eq:basic_total_var}
    \operatorname{Var}(z) = \mathbb{E}(\operatorname{Var}(z|W_n)) + \operatorname{Var}(\mathbb{E}(z|W_n))
\end{equation}
which decomposes the total variance into the variance within and between the scale-$n$ regions, respectively.  This decomposition is related to the spatial stratification of heterogeneity (see, e.g., the q-statistic developed by Wang et al. \cite{wang_measure_2016}). Recursively applying equation~\ref{eq:basic_total_var} to each of the two terms on its right-hand side and noting that $W_{i+1}$ is determined by $W_{i}$ (since smaller-scale regions are nested within larger-scale ones), we obtain
 \begin{align}
    \label{eq:variance_below}
    \mathbb{E}(\operatorname{Var}(z|W_n))=&\sum^{n-1}_{i=0}\mathbb{E}\left(\operatorname{Var}\left(\mathbb{E}\left(z \mid W_{i}\right) \mid W_{i+1}\right)\right)
     \\
    \operatorname{Var}(\mathbb{E}(z|W_n))=&\sum^N_{i=n}\mathbb{E}\left(\operatorname{Var}\left(\mathbb{E}\left(z \mid W_{i}\right) \mid W_{i+1}\right)\right) \label{eq:var_above} \\
            \operatorname{Var}(z)=&\sum^{N}_{i=0}\mathbb{E}\left(\operatorname{Var}\left(\mathbb{E}\left(z \mid W_{i}\right) \mid W_{i+1}\right)\right) \label{eq:total_var}
\end{align}
Equation~\ref{eq:total_var} is equivalent to equation~\ref{eqn:coarse_grain} of the main text. 
Here, we employ the notational shorthand $\mathbb{E}\left(\operatorname{Var}\left(\mathbb{E}\left(z \mid W_{0}\right) \mid W_{1}\right)\right) = \mathbb{E}\left(\operatorname{Var}\left(z \mid W_{1}\right)\right)$, since $W_0=z$, and $\mathbb{E}\left(\operatorname{Var}\left(\mathbb{E}\left(z \mid W_{N}\right) \mid W_{N+1}\right)\right) = \operatorname{Var}\left(\mathbb{E}\left(z \mid W_{N}\right)\right)$, since $W_{N+1}$ takes on only a single value (as it corresponds to the entire region in question). 
 
The terms on the right-hand sides of equations~\ref{eq:variance_below}-\ref{eq:total_var} correspond to the \textit{added} variance at scale $i+1$, while the left-hand side of equation~\ref{eq:var_above} corresponds to the total variance above scale $n$ (see figure~\ref{fig:precinct}).

Equation~\ref{eq:variance_below} denotes the minimum amount of variance that needs to be resolved on average by political decisions made at scale $n$ (regardless of whether such a decision is made through direct democracy, a single executive, or a legislature), since the smallest mean square distance achievable between the outcome and the electorate is the variance of the electorate opinions. Formally, 
\begin{equation}
    \int^{\infty}_{-\infty} (x - y)^2 f(x)dx \geq \int^{\infty}_{-\infty} (x - \mu)^2 f(x) dx
\end{equation}
for all outcomes $y$, since the mean $\mu$ of $f(x)$ minimizes $\int^\infty_{-\infty} (x - y)^2 f(x) dx$.
For instance, local elections in the U.S. need only resolve differences in opinions from within the locales, state elections must resolve both within-locale and within-state differences between locales, and national elections must resolve the total variance $\operatorname{Var}(z)$, which consists of within-locale, within-state, and between-state differences.  

While the analysis in this section has focused on one-dimensional random variables, it can easily be extended to multidimensional political opinions and outcomes by simply replacing the law of total variance with the law of total covariance.

\section{Effects of social ties on fully mixed and fully segregated electorates}
\label{appendix:network_structure}

We introduced the idea of effective opinions in section~\ref{sec:socialization}, wherein people vote as if they held a different political position because of their affiliation with their neighbors. This appendix explores how these  social ties influence political polarization when people consider the opinions of others in the electorate equally (the ``fully-connected" case) and when people's ties are determined purely by their political affiliation (the ``segregated" case). We then discuss an interpolation between the two extremes in appendix~\ref{appendix:geograpical_segregation} where disagreement and social ties possess a multiscale structure.

We first derive how the opinion distribution transforms under these ties. Generally, the effective opinion $x'$ of a voter can be written as some function of their opinion $x$. For monotone transformations $x' = t(x)$, the distribution of the random variable $x'$ in terms of $x$ is given by 
\begin{equation}
    \label{eq:transformation}
    f_{x'}(x') = f_{x}(t^{-1}(x'))|\frac{d}{dx'} t^{-1}(x')|.
\end{equation}

From this point on, we simply notate $f_{x'}(x')$ as $\hat{f}(x')$ and the original distribution as $f(x)$. Recall from equation~\ref{eq:social_opinion} that a citizen's effective opinion is defined as a weighted average of their own opinions and those of their neighbors, $x'_i = \sum_j T_{ij} x_j$. If we consider a connectivity matrix where a voter takes into account the opinions of every other member of the electorate equally, i.e. $T_{ij} = w/(n-1)$ for $i \neq j$ (where $n$ is the size of the electorate) while weighting their own position as $T_{ij} = 1 - w$ for $i = j$, we can write
\begin{equation}
\label{eq:w}
    x' = t(x) = x(1-w) - w\bar{x},
\end{equation}
for $n \gg 1$. Here, $w$ denotes the weight with which each person accounts for the opinion of others and $\bar{x}$ is the average of the opinion distribution. Using equation~\ref{eq:transformation}, we can write the transformed opinion distribution in terms of $f(x)$:
\begin{equation}
\label{eq:transformation_fully_connected}
    \hat{f}(x') = f\left(\frac{x' - w \bar{x}}{1-w}\right) \frac{1}{1-w}.
\end{equation}

The variance of the transformed distribution is then given by  
\begin{equation}
\label{eq:transformation_variance}
\begin{aligned}
    \hat{\sigma}^2 = \int^{\infty}_{-\infty} (x' - \bar{x})^2 f(\frac{x' - w \bar{x}}{1-w})\frac{1}{1-w} dx' \\ = \int^{\infty}_{-\infty} u^2(1-w)^2 f(u) du = \sigma^2 (1-w)^2,
\end{aligned}
\end{equation}
using $du = dx'/(1-w)$, where $\sigma^2 = \int^{\infty}_{-\infty}(x - \bar{x})^2 f(x)dx$ is the variance of $f(x)$. This is a general result for any $f(x)$: in a fully-connected locale, the variance of the opinion distribution decreases as the social weight $w$ is increased from $0$ to $1$. 

Next, we examine how social ties affect election stability in this fully-connected case. For simplicity (and because this distribution provides a precisely solvable case), we consider two normally distributed subpopulations of equal variance $\sigma^2$. This may describe two political parties with voters clustering around their respective means $\mu_A$ and $\mu_B$:
\begin{equation}
    f(x) = \pi_A e^{-\frac{(x-\mu_A)^2}{2\sigma^2}} + \pi_B e^{-\frac{(x-\mu_B)^2}{2\sigma^2}},
\end{equation}
where $\pi_A$ and $\pi_B$ are the relative sizes of the populations. For this distribution, 
\begin{equation}
J\equiv \frac{(\mu_A-\mu_B)^2}{4(\sigma^2+a^2)}
\end{equation}
for some positive constant $a$ gives a dimensionless measure of the degree of polarization.  Under a particular class of models, instability (see section~\ref{sec:single_dimensional}) occurs whenever $J>1$~\cite{siegenfeld_bar-yam_2020}; however, all of our arguments here will hold as long as a larger value of $J$ (which corresponds to a more hollowed-out center, relative to the length scale $a$) is more likely to produce instability.  

For social ties that result in an effective opinion distribution described by the transformation in equation~\ref{eq:w}, the distribution average,  $\bar{x} = (\pi_A \mu_A + \pi_B \mu_B)/(\pi_A + \pi_B)$, stays the same, while the means of the two subpopulations are shifted to $\hat{\mu}_A = \bar{x} w + \mu_A(1-w)$ and $\hat{\mu}_B = \bar{x} w + \mu_B(1-w)$, respectively, and their variances are decreased to $\hat{\sigma}^2 = \sigma^2(1-w)^2$.  The overall result of this transformation is to decrease the dimensionless polarization $J$ to  
\begin{equation}
\label{eq:instability_criterion}
    \hat J=\frac{(\mu_A - \mu_B)^2(1-w)^2}{4(\sigma^2(1-w)^2+a^2)}<J
\end{equation}
thus reducing the likelihood or magnitude of instability.

Having examined the case of a fully-connected locale, we now turn our attention to the situation where social ties are highly segregated.  In this limit, affective polarization arises where individuals sort their social interactions solely according to partisan affiliation. This corresponds to a graph with two disconnected components, where one component is fully populated by members whose opinion distribution is drawn from a Gaussian centered at $\mu_A$, and the other from the Gaussian at $\mu_B$. Each component is internally connected with weight $w$. Since this is just a sum of two fully-connected populations, we can apply the same transformation (equation~\ref{eq:transformation_fully_connected}) for each group. The effective means $\hat{\mu}_A$ and $\hat{\mu}_B$ stay the same because members of the two groups do not influence each other. However, because the widths of the two Gaussians  decrease to $\hat{\sigma} = \sigma (1-w)$, social ties may turn a stable election into an unstable one, since
\begin{equation}
    \hat J=\frac{(\mu_A - \mu_B)^2}{4(\sigma^2(1-w)^2+a^2)}>J.
\end{equation}
In contrast to the fully-connected case, increasing the strength of social ties can hollow out the middle, increasing rather than decreasing the possibility of instability in the election.

\section{Effects of social ties across multiple scales}
\label{appendix:geograpical_segregation}
Here we consider the effect of social ties in a multilevel setting, under the assumption that the strength of social ties is correlated with geographic proximity.    

We denote the opinion distribution of each region as $f_{s_1, s_2,\dots, s_N}(x)$, where $s_1, s_2, \dots, s_N$ are indices that specify the location of that region (e.g., $s_1$ may denote the city, $s_2$ the county, and $s_3$ the state). Dropping an index indicates an implicit sum: $f_{s_2,\dots, s_N}(x)$ is the total opinion distribution of the scale-$2$ region denoted by $s_2,...,s_N$, which is equal to the sum of the opinion distributions of all of the scale-$1$ regions it contains.  

We consider the simplest model that allows for the strength of social ties to vary with geographic scale, denoting the strength of social ties within scale-$n$ regions by $w_n$.  (For instance, in a three-scale model, $w_1$ could correspond to within-precinct ties, $w_2$ to within-county ties, and $w_3$ to state-wide ties.)  This model is more flexible than it may seem, since one can define arbitrarily many scales, allowing one to approach a continuum of possible strengths of social ties, with regions at each scale defined so as to give the desired strength of social ties between various groups of individuals (although of course this freedom is constrained by the nested structure of the regions at various scales).

Using a similar formulation as equation~\ref{eq:transformation}, we write the effective opinion of a voter residing in a region specified by $s_1, s_2, \dots, s_N$ as
\begin{equation}
    x' = x\beta + w_1 \bar{x}_{s_1,\dots,s_N} + \dots + w_N \bar{x}_{s_N} + w_{N+1}\bar x,
\end{equation}
where $\beta=1-\sum_{j=1}^{N+1} w_j$. The means follow the same summing notation, with $\bar{x}_{s_n,s_{n+1},...,s_N}$ being the mean of scale-$n$ region containing the voter ($\bar x$ denotes the average opinion of the entire nation). Since the effective transformed opinion of a region specified by $s_n,...,s_N$ is the sum over all the distributions within each all of the scale-$1$ regions it contains (each shifted by varying amounts since they are affected by social ties to different populations), we can write
\begin{equation}
\begin{aligned}
    \hat{f}_{s_n,...,s_N}(x') = \frac{1}{\beta}\sum_{s_{n-1} \in s_n}\dots\sum_{s_2 \in s_3}\sum_{s_1 \in s_2} \\ f_{s_1, \dots ,s_N}\left(\frac{x' -w_1 \bar{x}_{s_1,\dots,s_N} - w_2 \bar{x}_{s_2,\dots,s_N} \dots -w_{N+1}\bar x}{\beta}\right).
\end{aligned}
\end{equation}
The effective variance of opinions within locales at the smallest scale, specified by the distribution $\hat{f}_{s_1,...,s_N}(x')$, is reduced by the effects of ties across $w_1$ through $w_{N+1}$. However, the effective variance of the means $\bar{x}_{s_1,\dots,s_N}$ is reduced only through interactions from $w_2$ through $w_{N+1}$. This nested structure means that the effective variance at any particular level is only reduced by interactions across larger scales, yielding a multiscale distribution of effective variance of the form
\begin{equation}
\begin{aligned}
    \operatorname{Var}(z')= \ & \mathbb{E}\left(\operatorname{Var}\left(z \mid W_{1}\right)\right) (1-\Sigma^{N+1}_{i=1}w_i)^2 \\ & + \mathbb{E}\left(\operatorname{Var}\left(\mathbb{E}\left(z \mid W_{1}\right) \mid W_{2}\right)\right)(1-\Sigma^{N+1}_{i=2}w_i)^2 \\ &  + \dots + \operatorname{Var}\left(\mathbb{E}\left(z \mid W_{N}\right)\right) (1-w_{N+1})^2,
\end{aligned}
\end{equation}
employing a similar notation to equation~\ref{eqn:coarse_grain} where $z$ is a random variable that samples over voter opinions $x$ and $z'$ samples over effective opinions $x'$. This equation is equivalent to equation~\ref{eqn:xformedvar} in the main text. The degree to which social ties reduce the effective variance increases at smaller scales: the variance within locales, for instance, is decreased by ties among members of the locale and ties to others across the country. The variance of the locale means, on the other hand, is decreased only by cross-locale ties. When social ties are in play, what matters is not only the total variance of opinions in a country but \textit{how} that variance is distributed across scales. A country with more opinion differences at smaller (rather than larger) scales would have a lower effective total variance.

Akin to the analysis performed in appendix~\ref{appendix:network_structure}, we can also examine how social ties affect the stability of elections. For simplicity, we will consider only two scales, but similar results can be obtained for any number of scales.  

Consider, for instance, two states with the same total variance. In the first state, all counties have exactly the same opinion distribution, which we model as the sum of two normal distributions with means of $\mu_A=\Delta$ and $\mu_B=-\Delta$ (since without loss of generality, we can take the distributions to be centered at $0$) and each with variance $\sigma^2$. In the second state, all counties have unimodal opinion distributions, with half of the counties having opinion distributions centered at $\mu_A=\Delta$ and the other half at $\mu_B=-\Delta$.  

For the first state, where all counties are internally polarized but identical to each other, the effective opinion distribution is given by
\begin{equation}
    \hat{f}_{s_2}(x') = \frac{1}{\beta}\sum_{s_1\in s_2}f_{s_1,s_2}(x'/\beta),
\end{equation}
since $\bar x_{s_1,s_2}=\bar x_{s_2}=0$ for all counties $s_1$. This is akin to the case of a fully-connected, single-scale locale from the previous appendix: social ties decrease the dimensionless polarization $J$ to
\begin{equation}
    \hat J=\frac{\Delta^2\beta^2}{\sigma^2\beta^2+a^2}<J,    
\end{equation}
reducing the magnitude and likelihood of instability.

In the second state, the effective opinion distribution is given by
\begin{equation}
    \hat{f}_{s_2}(x') = \frac{1}{\beta}\sum_{s_1 \in s_2} f\left(\frac{x' - w_1 \bar{x}_{s_1, s_2}}{\beta}\right),
\end{equation}
since $\bar{x}_{s_2} = 0$ but the counties have different means $\bar{x}_{s_1, s_2}$ equal to either $\mu_A=\Delta$ or $\mu_B=-\Delta$. This transformation reduces the variance of each county by $\beta=1-w_1-w_2$, thus making the within-county populations more sharply peaked, but only changes the distance between the locales by $(1-w_2)$. Computing the effective $J$ yields
\begin{equation}
    \hat J=\frac{\Delta^2(1-w_2)^2}{\sigma^2(1-w_1-w_2)^2+a^2}.
\end{equation}
In this case, $\hat{J}$ is always equal to or larger than the $\hat{J}$ computed for first state, where all the polarization is concentrated at the smallest scale, leading to a higher chance of instability.

Increasing $w_1$ here increases $\hat{J}$, while increasing $w_2$ decreases $\hat{J}$; whether or not $\hat J$ is greater than or less than $J$ will depend on the precise values of the parameters. 
Intuitively, by decreasing local variance, stronger local social ties can result in a more hollowed-out center of the effective opinion distribution if there is substantial heterogeneity at larger scales, leading to a higher likelihood of larger-scale instability. Thus, for an electorate with substantial geographic segregation, the effect of social ties can be to decrease polarization and instability for local elections, while simultaneously increasing them for state and national elections.


We can frame this in the perspective of opinion sorting. We saw in the previous appendix that the segregation of opinions within a particular locale increases the degree of instability in an election. The multiscale model provides a more general result: segregation at any particular scale, where opinions in regions below that scale are homogenized and opinions above it are made more heterogeneous, can create strictly more instability for larger scale elections even when the overall opinion distribution of the system is held constant.

\section{Choice of election axes}
\label{appendix:candidate_selection}

\begin{figure*}
\includegraphics[width=1\textwidth]{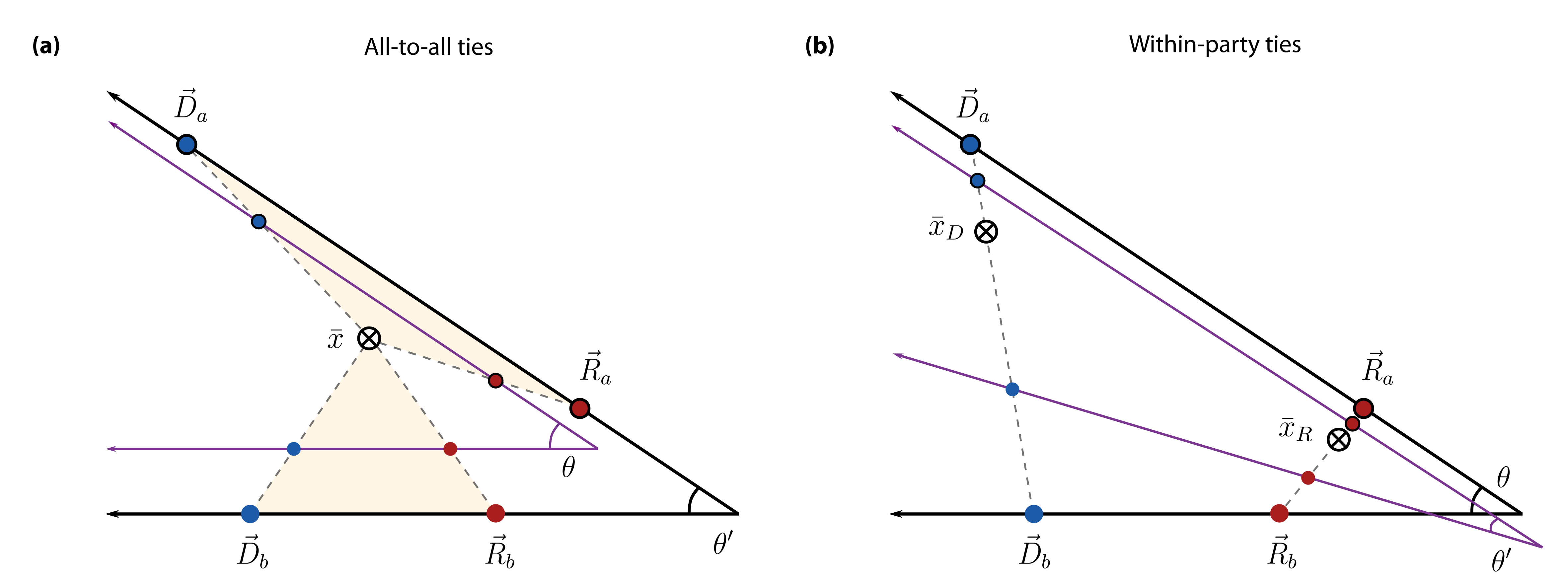}
\caption{\label{fig:axis_socialization} 
(a) When politicians are equally connected regardless of affiliation, the angle $\theta$ between axes $\hat{e}_a$ and $\hat{e}_b$ stays constant and the distances between the two parties ($|D'_{a}-R'_{a}|$ and $|D'_{b}-R'_{b}|$) decrease linearly with the social weight $m$. (b) On the other hand, under the assumptions stated in appendix~ \ref{appendix:axis_interaction}, if only intra-party ties exist, the angle $\theta'$ between the the axes decreases with $m$. 
}
\end{figure*}

Here we elaborate upon the concept of the election axis described in section~\ref{subsec:election_axes} and provide one model that links the election axis to the multidimensional opinion distribution of the electorate. The axis, which is simply the direction of discourse spanned by the eventual candidates in the general election, exists independently of any modeling assumptions, including those described below. 

We consider a specific model in which coalitions among potential voters (e.g. political parties or groups of parties) form so as to minimize of the ideological distance among potential voters within each coalition.   Letting $P=\{P_1,..,P_k\}$ be a partition of the set of all electorate opinions into $k$ coalitions, membership in each coalition is then given by 
\begin{equation}
    \arg\min_{P} \sum_{i = 1}^{k} \sum_{x \in P_i} |\vec{x} - \vec{\mu}_i|^2
\end{equation}
where $\vec{\mu}_i$ is the mean of $P_i$.  

For $k=2$, if opinions lie in the Euclidean space $\mathbb{R}^d$, the boundary between the two clusters is always given by a flat hyperplane of $\mathbb{R}^{d-1}$ since the Euclidean distance minimization corresponds to a linear kernel. We focus on the $k = 2$ case as it provides an analogue for the dominant social cleavage even in multi-party elections, although it is also possible to define a similar process for an arbitrary $k$ and a corresponding election subspace of dimension $k-1$.

Under these assumptions, the election axis is given by the difference in the means of the two clusters $\vec{\mu}_1 - \vec{\mu}_2$. A useful heuristic is to think of this axis as an approximation to the the first principal component $v_1$ of the opinion distribution $f(\vec{x})$: as shown by Ding and He \cite{ding_he}, the first principal component of the opinion distribution $f(\vec{x})$ approximates  the continuous solutions to the discrete cluster membership indicators for 2-means clustering. Even though the categorical constraint of the two clusters means that the centroid vector and the first principal component do not coincide exactly, we can employ $v_1$ as the leading order proxy for the election axis. 

\section{Election axis interactions}
\label{appendix:multilevel_axis}

Elections in a democracy interact with each other via a variety of mechanisms, such as partisan ties, shared electorates, and media-driven alignment. In this section, we develop a general framework for how two election axes, specified by unit vectors $\hat{e}_a$ and $\hat{e}_b$, may couple to each other. We then show in the following appendix that this formulation is compatible with the idea of effective opinions (as introduced in section~\ref{sec:socialization}).

We can describe the interaction of two elections by writing their new axes $\hat{e}'_a$ and $\hat{e}'_b$ as a linear combination of the original directions:
\begin{equation}
\label{eq:linear_combination}
\begin{split}
    \hat{e}'_a = \frac{w_a \hat{e}_a + (1-w_a) \hat{e}_b}{|w_a \hat{e}_a + (1-w_a) \hat{e}_b|} \\
    \hat{e}'_b = \frac{w_b \hat{e}_b + (1-w_b) \hat{e}_a}{|w_b \hat{e}_b + (1-w_b) \hat{e}_a|},
\end{split}
\end{equation}
where $w_a,w_b\in [0,1]$. A combination of this form assumes that the coupled elections lie within the span of the original axes. In other words, it assumes that no new political issues are produced by the interaction; the old issue dimensions are merely mixed. The relationship between elections $a$ and $b$ can be asymmetrical: they may be conducted at different levels (e.g., one is national and another is local), one election may be more salient than the other, or they may simply involve electorates with different populations. 

To see the effect of increasing the interaction strengths $w_a$ and $w_b$, note that we can always pick a coordinate system for $\hat{e}_a$ and $\hat{e}_b$ such that they span a plane, where the initial angle between them is given by some $\theta_0$. If we pick a coordinate system where $\hat{e}_a = (1, 0 ,0 \ldots)$ and $\hat{e}_b = (\cos\theta_0, \sin\theta_0, 0 \ldots)$, we see that we can write $\hat e_a'= (\cos\theta_a, \sin\theta_a, 0 \ldots)$ and $\hat e_b'= (\cos\theta_b, \sin\theta_b, 0 \ldots)$ for some $\theta_a, \theta_b\in [0,\theta_0]$.  Since $\theta_a$ increases with increasing $w_a$ and $\theta_b$ decreases with increasing $w_b$, the angle between the two axes $\theta_b-\theta_a$ decreases with stronger interactions (i.e. larger values of $w_a$ and/or $w_b$).

We can extend this two-axis model to a country with elections at scales $\alpha, \beta, \gamma \ldots$. In the most general setting, a multilevel system can have both horizontal interactions (e.g., those driven by shared party mobilization between states) and vertical interactions (e.g., those driven by the effect of a shared electorate in nested elections). Each scale consist of sets of axes $\boldsymbol{\alpha} = \{\alpha_1, \alpha_2, \ldots\},$ $\boldsymbol{\beta} = \{\beta_1, \beta_2 \ldots\}$, etc. The un-normalized effective axis for an election $\hat{e}_i$---produced as a result of multilevel interactions---can be written as 
\begin{equation}
    e_i' = w\hat{e}_i + (1-w)\left(\sum_{j, \hat{\alpha}_j \neq \hat{e}_i}^{|\boldsymbol{\alpha}|} A_{ij} \hat{\alpha}_j + \sum_{j, \hat{\beta}_j \neq \hat{e}_i}^{|\boldsymbol{\beta}|} B_{ij} \hat{\beta}_j  + \cdots \right),
\end{equation}
where $A_{ij}, B_{ij}, \dots$ represent interaction matrices with elections at the respective scales and $w$ parameterizes the total strength of these interactions. The normalized axis is given by $\hat{e}'_i =  e_i'/| e_i'|.$ A version of this model with two scales is described in the main text.

Generalizing the pairwise interaction shown in equation~\ref{eq:linear_combination}, the multilevel interactions help to focus all the election axes in a country---whose discourse may originally be oriented along any number of issue dimensions---toward a single direction. For instance, we can measure the dispersion of the axes $\hat{\mathbf{e}}_{i}$ via the circular variance of the subtended angle $\theta' = \operatorname{cos}^{-1}(\hat{\mathbf{e}}_{i} \cdot \hat{\mathbf{e}}_{N})$,
\begin{equation}
    \operatorname{Var}(\theta') = 1 - \frac{1}{n}\sqrt{\sum^n_{i = 1} \cos^2(\theta'_i) + \sum^n_{i = 1}\sin^2(\theta'_i)},
\end{equation}
where $i$ indexes through all the elections in the country. This decreases monotonically with stronger social ties. Although the focusing effect can, in principle, lead to greater coherence in a country's legislature, it also collapses the dimensionality of its political discourse. As described in section~\ref{sec:madison}, the lower dimensionality tends to increase the projected variance along the first principal axis of legislator opinions, an effect that can be measured via DW-NOMINATE scores.

\section{Interactions via partisan ties}
\label{appendix:axis_interaction}

Here, we explore a specific model of how elections interact based on the effective opinion model in section~\ref{sec:socialization} to provide an example of how interactions between election axes may arise. We show that this model is consistent with the more general coarse-grained formulation of axis interactions introduced above in appendix~\ref{appendix:multilevel_axis}. Other specific models will also be consistent with the coarse-grained formulation. While real-world elections will of course not follow the precise dynamics given here, they may nonetheless be well described in aggregate by the coarse-grained formulation. 

Take, as an example, elections $a$ and $b$ with Democratic and Republican candidates located at $\vec{D}_{a}$, $\vec{R}_{a}$, and $\vec{D}_{b}$, $\vec{R}_{b}$ respectively. The candidates span election axes $\hat{e}_i = (\vec{D}_{i} - \vec{R}_{i})/|\vec{D}_{i} - \vec{R}_{i}|$, where $i \in\{a, b\}$. Equation~\ref{eq:transformation} tells us that in order to compute the transformed positions when all opinion holders are equally connected, we find the mean $\bar{x}$ of the distribution and shift the original opinions proportional to a weight $m \in [0, 1]$. If all the politicians in a country are equally connected, the new position of the Democratic candidate in election $i$ can be written as,
\begin{equation}
    \vec{D}'_{i} = \bar{x} m + \vec{D}_{i}(1-m).
\end{equation}
The expressions for the Republican candidate follows identically. As a result, the distance between candidates of opposing parties,
\begin{equation}
    |\vec{D}'_{i} - \vec{R}'_{i}| = (1-m)|\vec{D}_{i} - \vec{R}_{i}|,
\end{equation}
decreases as the strength of social ties is increased. Furthermore, the angle $\theta' = \operatorname{cos}^{-1}(\hat{e}_a' \cdot \hat{e}_b')$ stays the same. This can be seen in figure~\ref{fig:axis_socialization}a: because $\vec{D}_{i}$ and $\vec{R}_{i}$ are both shifted toward the center of mass by the same proportion, they form a pair of similar triangles with the transformed candidate positions. As a result, the axes are always translated parallel to their original directions.

However, we obtain a different result if interactions that result in changes in effective political opinions exist predominantly within (rather than between) parties. When social ties only occur in-party, we let the effective opinions of Democrats  move towards $\bar{x}_{D} = p_{a} \vec{D}_{a} + p_{b} \vec{D}_{b}$ and those of Republicans move towards $\bar{x}_{R} = p_{a} \vec{R}_{a} + p_{b} \vec{R}_{b}$, with $p_{a}$ and $p_{b}$ parameterizing the relative size or salience of elections $a$ and $b$ (and with $p_a + p_{b}= 1$). If effective opinions are pulled toward these means with weight $m$, the new election axes become $\hat{e}'_i = \vec{e}'_i/|\vec{e}'_i|$, where
\begin{equation}
\begin{aligned}
    \vec{e}'_a \ = (\bar{x}_{D}-\bar{x}_{R}) m + (\vec{D}_{a} - \vec{R}_{a})(1-m) \\ = (\vec{D}_{a}-\vec{R}_{a})(1-m+p_{a}m) + (\vec{D}_{b}-\vec{R}_{b})p_{b}m.
\end{aligned}
\end{equation}
A similar expression can be obtained for $\vec{e}'_b$.  Mapping $m - p_{a}m = p_{b}m$ onto $w$, we see that this is a linear combination of $\vec{D}_{a}-\vec{R}_{a}$ and $\vec{D}_{b}-\vec{R}_{b}$ of the form presented in equation~\ref{eq:linear_combination}. As we increase the level of within-party socialization $m$, the angle between the two elections axes $\hat{e}'_a$ and $\hat{e}'_b$ decreases.

\section{Multidimensional representation}
\label{appendix:multidimensional_representation}

Recall from section~\ref{sec:single_dimensional} that the representation $r_i$ of an individual $i$ is defined as the effect of a shift in their opinion $x_i$ on the election outcome. While representation is either positive or negative in the one-dimensional case, in a multidimensional opinion space, the outcome of an election can change in any direction relative to the change in $x_i$.

This requires us to generalize the representation $r_i$ as a tensor, taking into account the direction in which the outcome changes for a given change in opinion. Assuming the derivative exists, we write the representation of opinion $x_i$ as
\begin{equation}
    r^i_{\mu\nu} = \frac{\partial y_\mu}{\partial x^i_\nu},
\end{equation}
where the second index corresponds to the direction of opinion change and the first index corresponds to the direction of change in the election outcome. This is a rank-two tensor, but because the opinion space can be embedded in $\mathbb{R}^n$, the metric is simply $\delta_{\mu\nu}$. Working in an Euclidean space enables us to lower all indices for notational simplicity. A similar notion of multidimensional representation was first presented in the first supplemental section of \cite{siegenfeld_bar-yam_2020}. 

A change in opinion in direction $\hat{c}$ can always be broken down into a component along the election axis $\hat{e}$ and a component along some orthogonal axis $\hat{o}$:
\begin{equation}
    \hat{c} = a \hat{e} + b \hat{o}.
\end{equation}
The differential representation along $\hat{c}$, i.e., the change in the election outcome along  $\hat{c}$ for a change in opinion in the same direction, can be written as
\begin{equation}
    \label{eq:multidimensional_repn}
    \begin{aligned}
    r_{\hat{c}} = & \hat{c}_{\mu} r^i_{\mu\nu} \hat{c}_{\nu} = \\ & a^2 \hat{e}_{\mu} r^i_{\mu\nu} \hat{e}_{\nu} + b^2 \hat{o}_{\mu} r^i_{\mu\nu} \hat{o}_{\nu} + ab(\hat{e}_{\mu} r^i_{\mu\nu} \hat{o}_{\nu} + \hat{o}_{\mu} r^i_{\mu\nu} \hat{e}_{\nu}).
    \end{aligned}
\end{equation}
The final term, which represents the change in the election outcome orthogonal to the change in opinion, vanishes when $\hat{e}$ is an eigenvector of $r^i_{\mu\nu}$. 

In general, equation~\ref{eq:multidimensional_repn} allows us to write the total representation as the sum of contributions from on-axis changes in opinion $a^2 \hat{e}_{\mu} r^i_{\mu\nu} \hat{e}_{\nu} + ab \hat{o}_{\mu} r^i_{\mu\nu} \hat{e}_{\nu}$ and off-axis changes in opinion $ b^2 \hat{o}_{\mu} r^i_{\mu\nu} \hat{o}_{\nu} + ab \hat{e}_{\mu} r^i_{\mu\nu} \hat{o}_{\nu}$. These components may have very different properties depending on the election process. For instance, if negative representation occurs strongly along the election axis---perhaps due to its correlation with national political discourse---changes in opinion along orthogonal directions may provide a good avenue for political representation. However, there may also be situations in which only changes in opinion along the direction of the axis are represented.  For instance in a local election dominated by national discourse, local issues that do not fall along the direction of national discourse may have little effect on the election outcome.

\end{document}